\documentclass[onecolumn,
 showpacs,amsmath,amssymb,11pt]{revtex4}
\usepackage{graphicx}
\usepackage{epstopdf}
\usepackage{bm}
\usepackage{subfigure}
\usepackage[latin1]{inputenc}
\usepackage{rotating}
\usepackage{longtable}
\usepackage{float}
\usepackage[normalem]{ulem}
\usepackage{dcolumn}
\usepackage{bm}
\usepackage{color}
\usepackage{footnote}

\usepackage{color, soul}
\usepackage{braket}

\begin{document}

\preprint{APS/123-QED}

\title{A high-performance Raman-Ramsey Cs vapor cell atomic clock}
\author{Moustafa Abdel Hafiz$^1$, Gr\'egoire Coget$^1$, Peter Yun$^2$, St\'ephane Gu\'erandel$^2$, Emeric de Clercq$^2$, Rodolphe Boudot$^1$}

\affiliation{$^1$FEMTO-ST, CNRS, UBFC, 26 chemin de l'Epitaphe 25030 Besan\c{c}on Cedex, France}
\affiliation{$^2$LNE-SYRTE, Observatoire de Paris, PSL Research University, CNRS, Sorbonne Universit\'es, UPMC Univ. Paris 06, 61 avenue de l'Observatoire, 75014 Paris, France}

\date{\today}

\begin{abstract}
We demonstrate a high-performance coherent-population-trapping (CPT) Cs vapor cell atomic clock using the push-pull optical pumping technique (PPOP) in the pulsed regime, allowing the detection of high-contrast and narrow Ramsey-CPT fringes. The impact of several experimental parameters onto the clock resonance and short-term fractional frequency stability, including the laser power, the cell temperature and the Ramsey sequence parameters, has been investigated. We observe and explain the existence of a slight dependence on laser power of the central Ramsey-CPT fringe line-width in the pulsed regime. We report also that the central fringe line-width is commonly narrower than the expected Ramsey line-width given by $1/(2T_R)$, with $T_R$ the free-evolution time, for short values of $T_R$. The clock demonstrates a short-term fractional frequency stability at the level of $2.3 \times 10^{-13}~\tau^{-1/2}$ up to 100 seconds averaging time, mainly limited by the laser AM noise. Comparable performances are obtained in the conventional continuous (CW) regime, if use of an additional laser power stabilization setup. The pulsed interaction allows to reduce significantly the clock frequency sensitivity to laser power variations, especially for high values of $T_R$. This pulsed CPT clock, ranking among the best microwave vapor cell atomic frequency standards, could find applications in telecommunication, instrumentation, defense or satellite-based navigation systems.
\end{abstract}

\maketitle
\clearpage
\section{Introduction}\label{sec:amx-introduction}
Numerous industrial applications including satellite-based navigation, telecommunication and space applications require field-deployable atomic clocks combining excellent fractional frequency stability, low power consumption and small size \cite{GPS, Galileo, Glonass, Chine:RSI:2016}. In this domain, lamp-based microwave Rb vapor cell atomic clocks based on optical-microwave double resonance technique, have been widely used for decades. The advent of high-performance narrow-line semiconductor diode lasers has later allowed to improve by at least one order of magnitude the stability performances of these clocks \cite{Camparo:JAP:1986, Mileti:IEEE:1998, Vanier:APB:2007}. Over the last years, relevant efforts have been pursued in different laboratories to push to the limit vapor cell clocks performances by adopting original schemes and approaches including mainly pulsed optical pumping (POP) \cite{Micalizio:Metrologia:2012} and coherent population trapping (CPT) \cite{Alzetta:PM:1976, Vanier:APB:2005} often associated with a pulsed Ramsey scheme. In this domain, INRIM has developed a pulsed optically pumped (POP) Rb frequency standard which exhibits a fractional frequency stability of $1.7 \times 10^{-13}~\tau^{-1/2}$ up to 1000 s integration time \cite{Micalizio:Metrologia:2012}. The POP technique is also studied and investigated in other groups \cite{Lin:OL:2012, Kang:JAP:2015}, demonstrating promising results. LNE-SYRTE has developed a high-performance Cs CPT clock, combining a pulsed Ramsey-like interrogation scheme and an optimized CPT pumping scheme called lin $\perp$ lin \cite{Zanon:PRL:2005}. This Cs CPT clock has demonstrated an Allan deviation of $3.2 \times 10^{-13}~\tau^{-1/2}$ up to averaging times of 100 s \cite{Danet:UFFC:2014}. However, this table-top clock prototype, using two external cavity diode lasers phase-locked with an optical phase lock loop, remains complex, voluminous and not-well adapted for integration and potential future industrial transfer. Moreover, the use of two phase-locked lasers degrades the microwave phase noise, enhancing the Dick effect contribution \cite{Dick:PTTI:1987} to the clock short-term frequency stability. Another promising and simple-architecture CPT clock developed in LNE-SYRTE, based on an original constructive polarization modulation technique \cite{Yun:APL:2014, Yun:JAP:2016}, has recently demonstrated an Allan deviation at the level of $3 \times 10^{-13}~ \tau^{-1/2}$ up to 100 s in the continuous-regime \cite{Yun:PRAp:2017}.\\
In the frame of the MClocks project \cite{Mclocks}, we have recently reported the development of a Cs CPT atomic clock based on a single-modulated laser source using the push-pull optical pumping (PPOP) technique \cite{Jau:PRL:2004, Liu:PRA:2013}. This clock, operating in the continuous (CW)-regime, has demonstrated a short-term fractional frequency stability of $3 \times 10^{-13} \tau^{-1/2}$ up to 100 s averaging time \cite{MAH:JAP:2015}. Nevertheless, it was reported that performances of this clock were strongly limited by laser power-induced frequency-shift effects. In that sense, we investigate in the present paper the application of a pulsed Ramsey-like interrogation scheme to this clock. The pulsed regime allows to detect high-signal-to-noise ratio and narrow Ramsey-CPT fringes and to reduce the sensitivity of the clock frequency to laser power variations \cite{Castagna:UFFC:2009, Boudot:IM:2009}. A detailed investigation is reported on the impact of the laser power, the cell temperature and the Ramsey sequence parameters on the central Ramsey fringe properties and short-term fractional frequency stability. This study has led to the demonstration of a clock short-term fractional frequency stability at the level of $2.3 \times 10^{-13} \tau^{-1/2}$ up to 100 seconds averaging time. Comparable performances are obtained in the CW regime, if use of an additional laser power stabilization setup. The sensitivity of the clock frequency to the laser power variations is reported for several values of the free-evolution time $T_R$, demonstrating a relevant reduction of this effect compared to the CW-regime case. These preliminary results are promising for the development of a high-performance CPT clock with improved mid-term and long-term fractional frequency stability.

\section{Experimental set-up}\label{sec:setup}
\begin{figure*}[t]
\centering
\includegraphics[width=\linewidth]{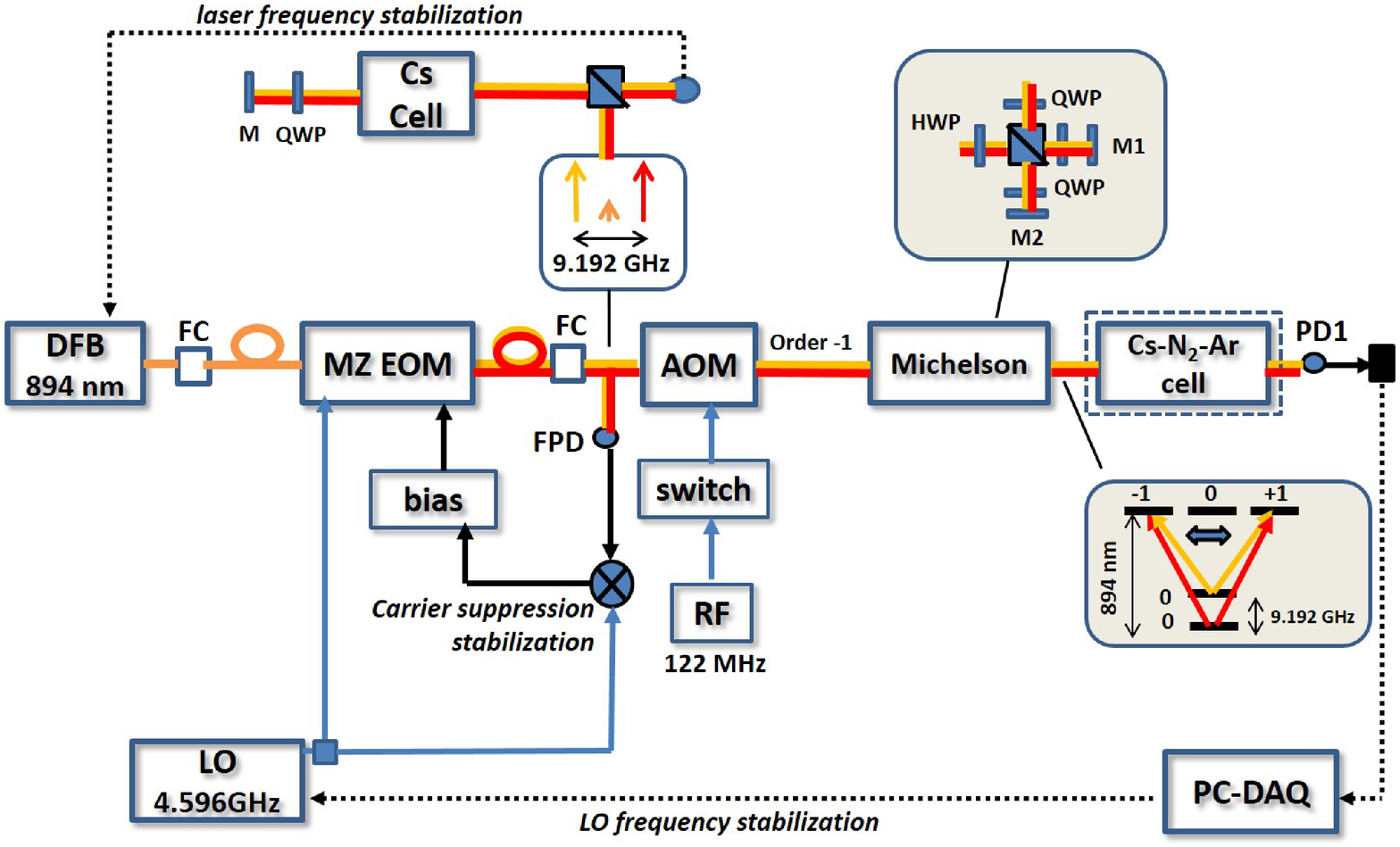}
\caption{}
\label{fig:setup-2}
\end{figure*}

\begin{figure*}[t]
\centering
\subfigure[]{\includegraphics[width=\linewidth]{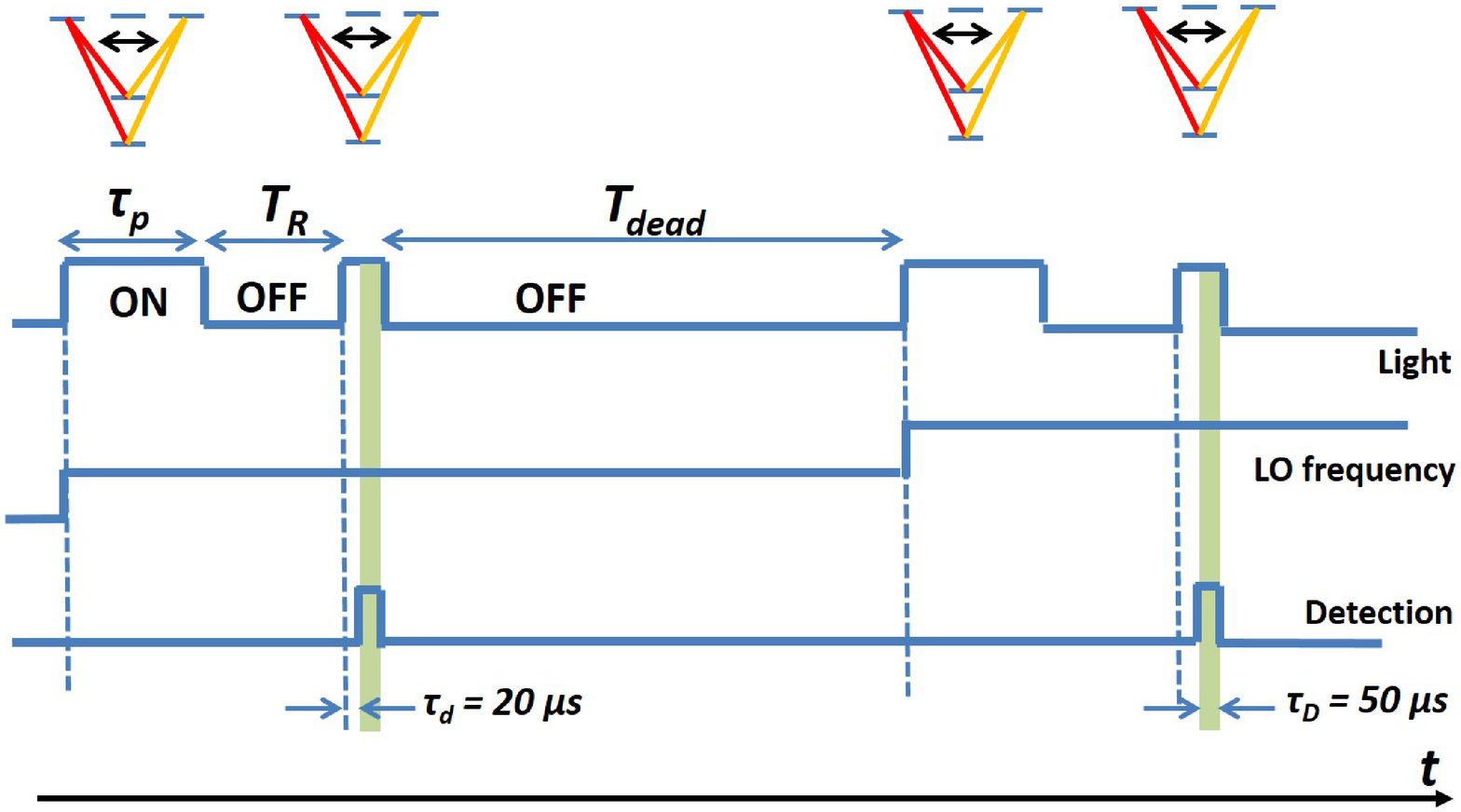}
\label{fig:spectro}} \vfill
\subfigure[]{\includegraphics[width=\linewidth]{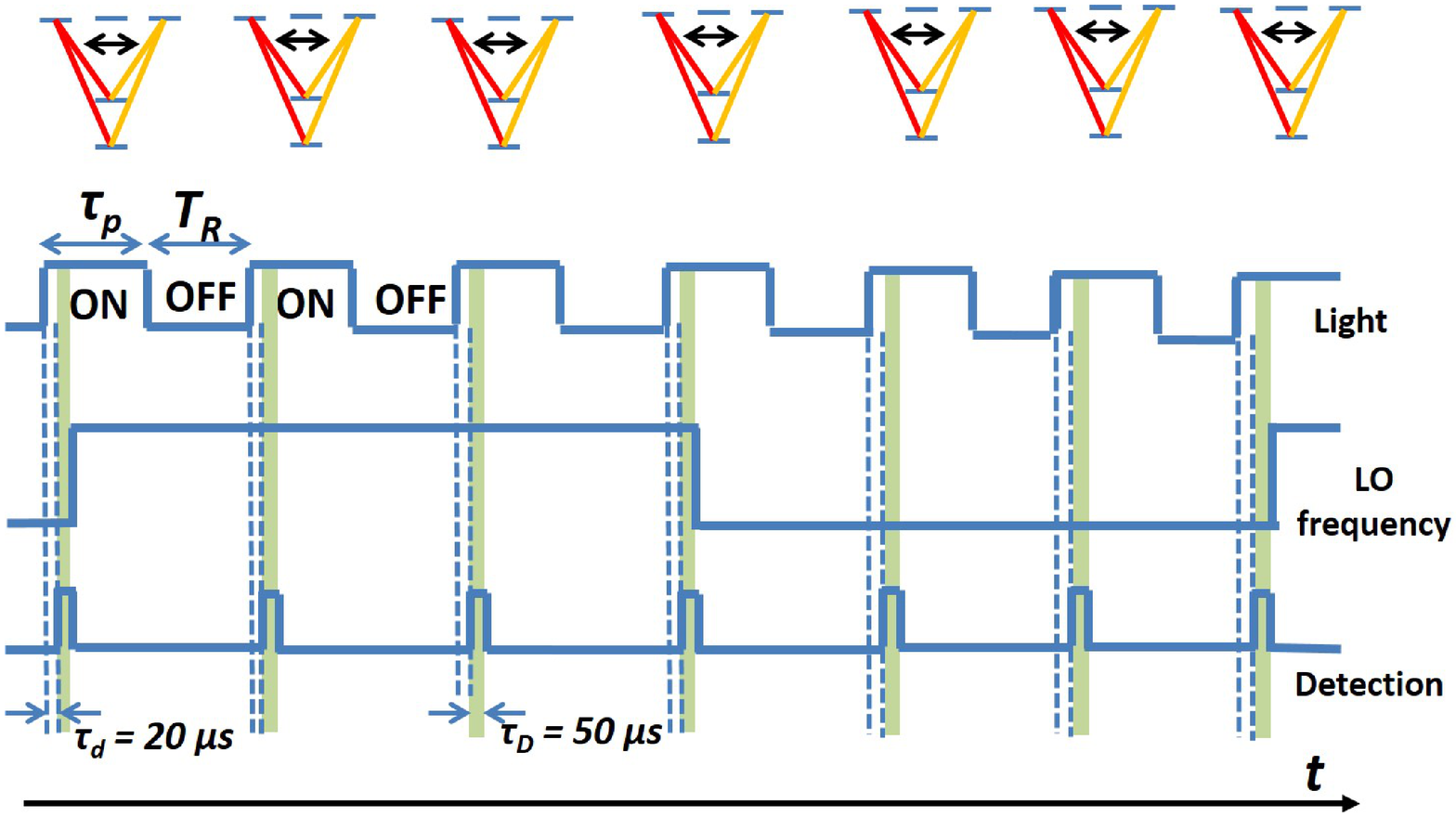}
\label{fig:lock}}
\caption{}
\end{figure*}

Figure \ref{fig:setup-2} presents the Cs CPT clock set-up. The laser source is a 1-MHz linewidth Distributed Feedback (DFB) diode laser tuned on the Cs D$_1$ line at 894.6 nm. Two optical isolation stages (not shown on Fig. \ref{fig:setup-2}), with a total isolation of about 70 dB, are placed at the output of the DFB laser to prevent optical feedback. First-order optical sidebands frequency-split by 9.192 GHz are generated by driving a pigtailed Mach-Zehnder electro-optic modulator (EOM - Photline NIR-MX800-LN-10) at 4.596 GHz with a low noise microwave frequency synthesizer \cite{Francois:RSI:2014}. At the output of the EOM, active optical carrier suppression stabilization is performed thanks to an original microwave synchronous detector \cite{Liu:PRA:2013}. The EOM is actively temperature-stabilized at about 40.2$^{\circ}$C where the optical power transmission is maximized and where the EOM transmission sensitivity to temperature variations is canceled at the first order \cite{MAH:JAP:2015}. At the output of the EOM, a fraction of the laser power is sent into a Fabry-Perot interferometer to analyze the optical power spectrum. Another fraction of the power is extracted to be sent in a dual-frequency Doppler-free spectroscopy setup for laser frequency stabilization \cite{MAH:OL:2016}. The laser fractional frequency stability is measured to be better than $2 \times 10^{-12}$ at 1 s integration time. At the direct output of the EOM, an acousto-optic modulator (AOM) is used. Its first role is to shift the laser frequency by $-$122 MHz to compensate for the buffer-gas induced optical frequency shift \cite{Pitz:PRA:2009} in the CPT cell. The second function of the AOM is to apply the Ramsey-like pulsed interrogation scheme. A microwave switch is used to turn on and off the AOM radio-frequency (RF) driving signal, allowing to switch on and off the first-order diffracted light beam of interest. Note that operating the clock in the CW regime, this AOM can also be used for laser power stabilization \cite{MAH:JAP:2015}. In the present study, no laser power stabilization was applied in the pulsed regime. Following the AOM, a Michelson delay-line and polarization orthogonalizer system, based on two arms each composed of a mirror and a quarter-wave plate, is used to generate two time-delayed orthogonally polarized optical fields. A last quarter wave plate allows to produce a bi-chromatic optical field that alternates between right and left circular polarization in order to realize the PPOP interaction scheme. The latter helps to  maximize the number of atoms into the magnetic-field insensitive 0-0 clock transition \cite{Jau:PRL:2004}. At the output of the Michelson system, the diameter beam is expanded to 2 cm thanks to a pair of convergent optical lenses to cover the whole diameter of the CPT cell. The latter is a 2-cm diameter and 5-cm long Cs vapor cell, filled with a N$_2$-Ar buffer gas mixture of total pressure 15.3 Torr and partial pressure ratio $r = P_{Ar}/P_{N_2}$ = 0.6. The cell is maintained into a temperature-stabilized copper oven. For most of the results reported in this paper, the cell temperature $T_{cell}$ is 35$^{\circ}$C. A solenoid is used to apply a static magnetic field ($B$ = 4.5 $\mu$T) parallel to the laser beam propagation. The ensemble is surrounded by a two-layer mu-metal magnetic shield. In the following, the cell input and output laser power are noted $P_i$ and $P_o$ respectively. CPT spectroscopy is performed by detecting the laser power transmitted through the CPT cell using the low-noise photodiode PD1. A data acquisition card-based automation platform, piloted by a computer with a dedicated software, allows to monitor and pilot numerous experimental parameters and to manage a large number of servo loops for proper and routine operation of the clock.\\
Two different pulses sequences are used in this study. The typical sequence used to perform Raman-Ramsey spectroscopy is shown on Fig. \ref{fig:spectro}. Atoms interact with optical CPT pulses. A first pulse of length $\tau_p$ allows to pump the atoms into the CPT state. Atoms then evolve freely in the dark during a time $T_R$. A second pulse is used for CPT signal detection. A delay $\tau_d$ of 20 $\mu$s is taken before opening a detection window with a length $\tau_D$. The clock signal is averaged over the detection window with a rate of 1 Msamples/s. A significant dead-time of 35 ms ($\sim 10~T_2$, with $T_2$ the hyperfine coherence relaxation time) is applied between each data point measurement to ensure that most of the atoms relaxed from the CPT state before entering in the next cycle. The sequence used in clock closed loop operation is shown on Fig. \ref{fig:lock}. In this regime, atoms interact with a CPT pulse sequence where each pulse is used both for CPT pumping and CPT detection. The sequence cycle duration is noted $T_c = \tau_p + T_R$. The associated cycle frequency is noted $f_c = 1/T_c$. The local oscillator (LO) frequency is changed every 3 cycles just after the detection window. The clock output signal is compared to the signal of a state-of-the-art reference active hydrogen maser with a fractional frequency stability of 8 $\times$ 10$^{-14}$ and 3 $\times$ 10$^{-15}$ at $\tau$ = 1 s and 100 s respectively  \cite{T4science}.
\section{Experimental results}\label{sec:results}
\subsection{Short-term stability}\label{sec:sts}
\subsubsection{Laser power $P_i$ and time sequence parameters} \label{sec:pt}
\begin{figure*}[t]
\centering
\includegraphics[width=\linewidth]{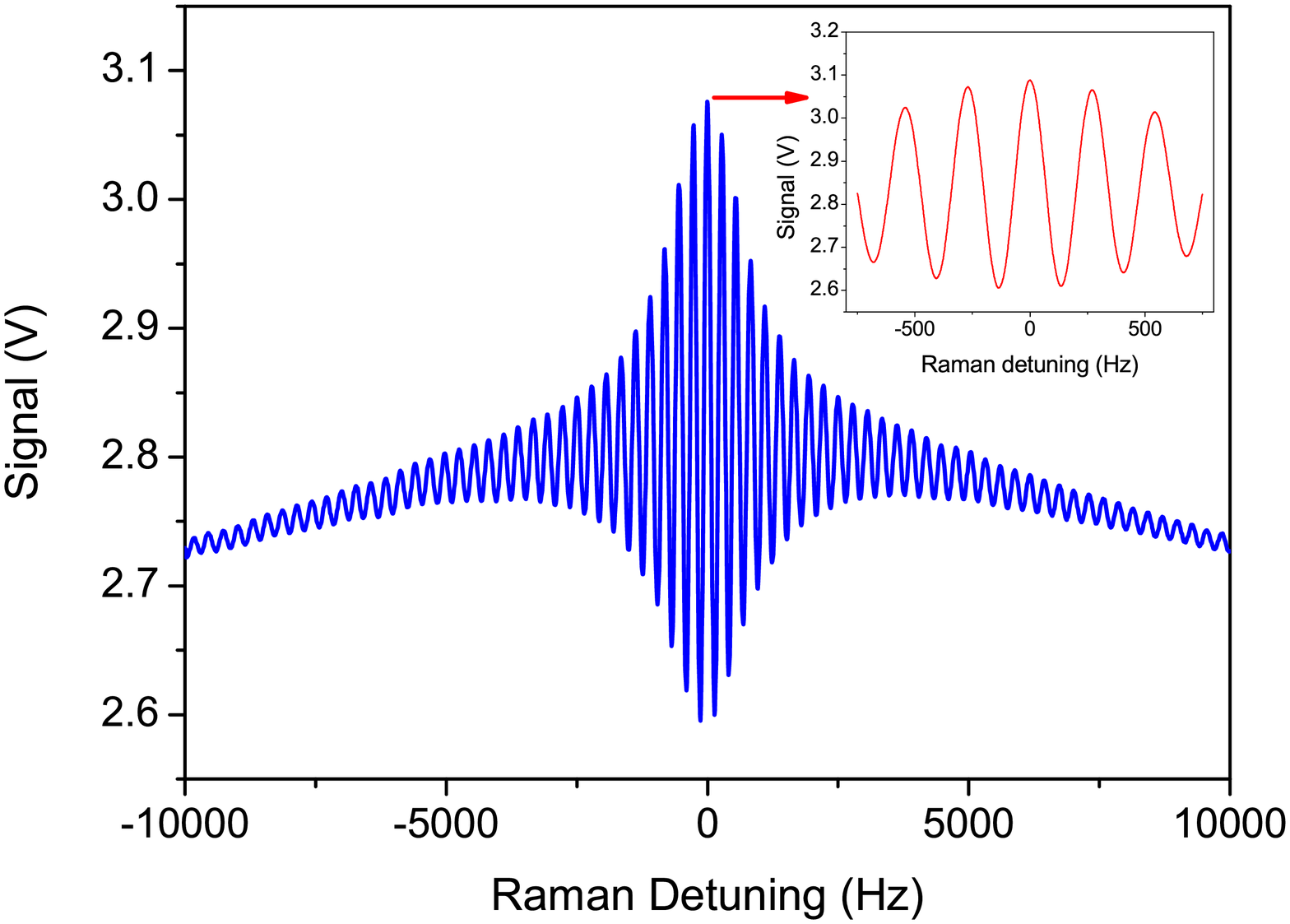}
\caption{}
\label{fig:fringe}
\end{figure*}

\begin{figure*}[t]
\centering
\subfigure[]{\includegraphics[width=0.8\linewidth]{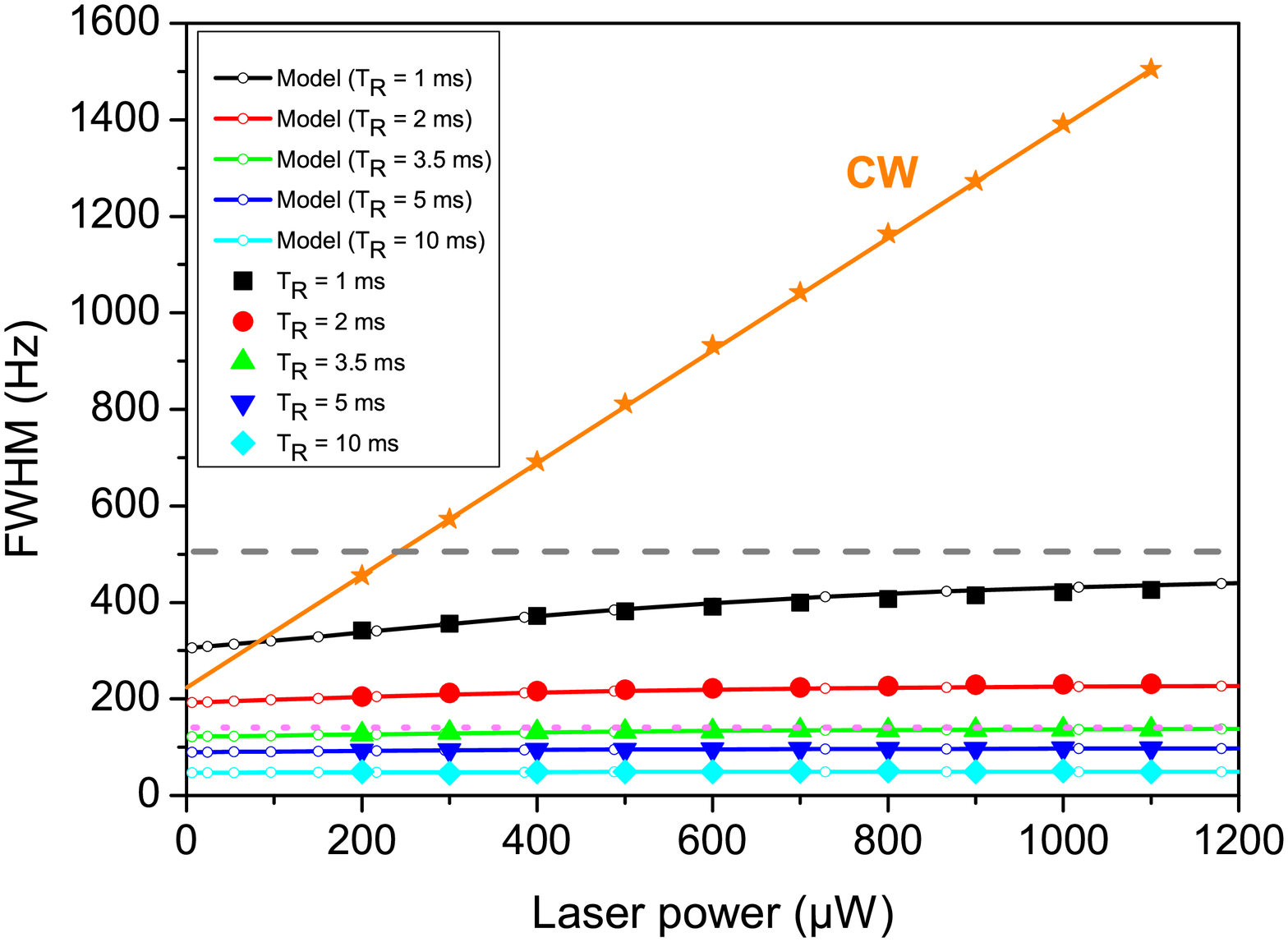}
\label{fig:fwhm-cw-pulsed}} \vfill
\subfigure[]{\includegraphics[width=0.8\linewidth]{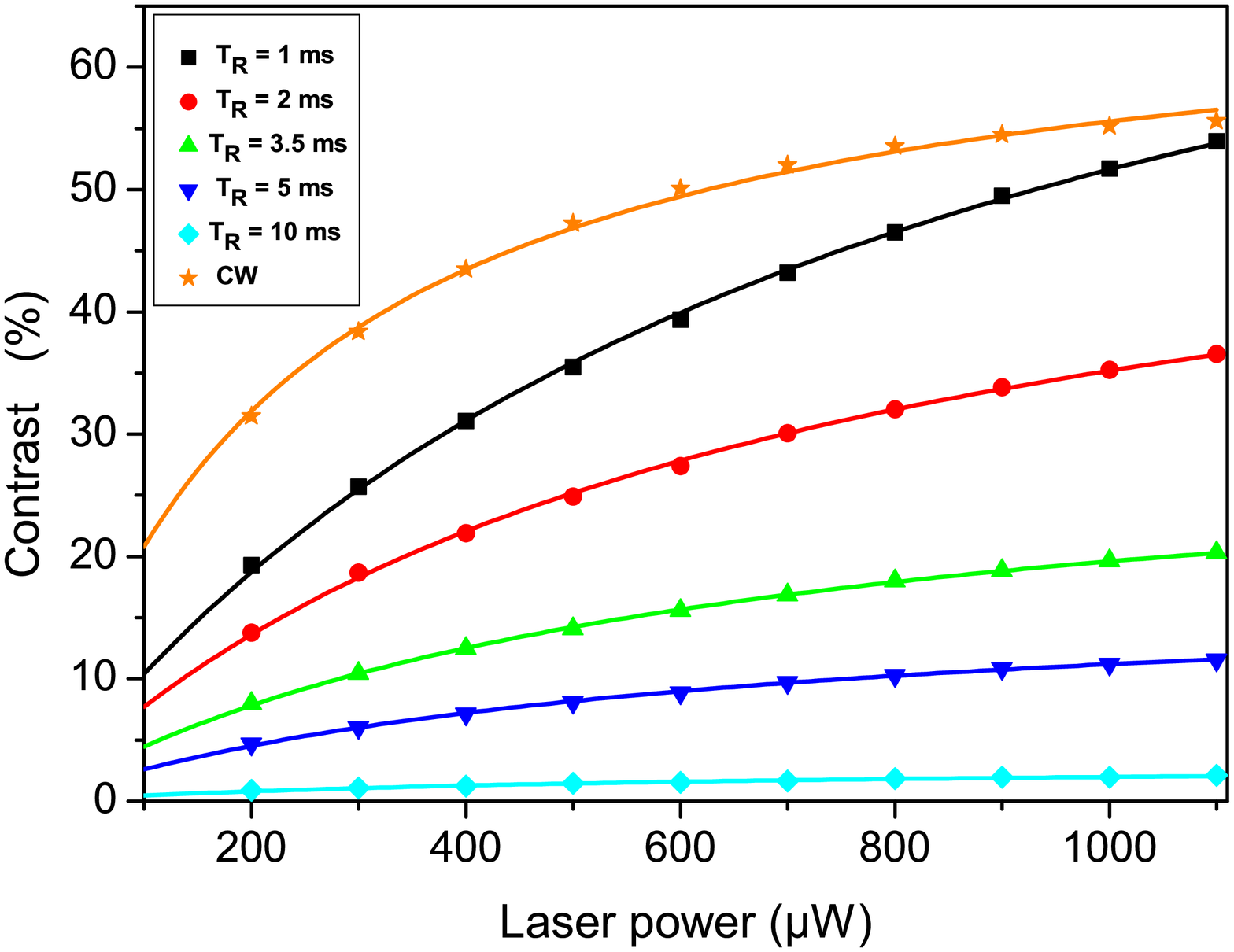}
\label{fig:c-p-TR}}
\caption{}
\end{figure*}

\begin{figure*}[t]
\centering
\subfigure[]{\includegraphics[width=0.5\linewidth]{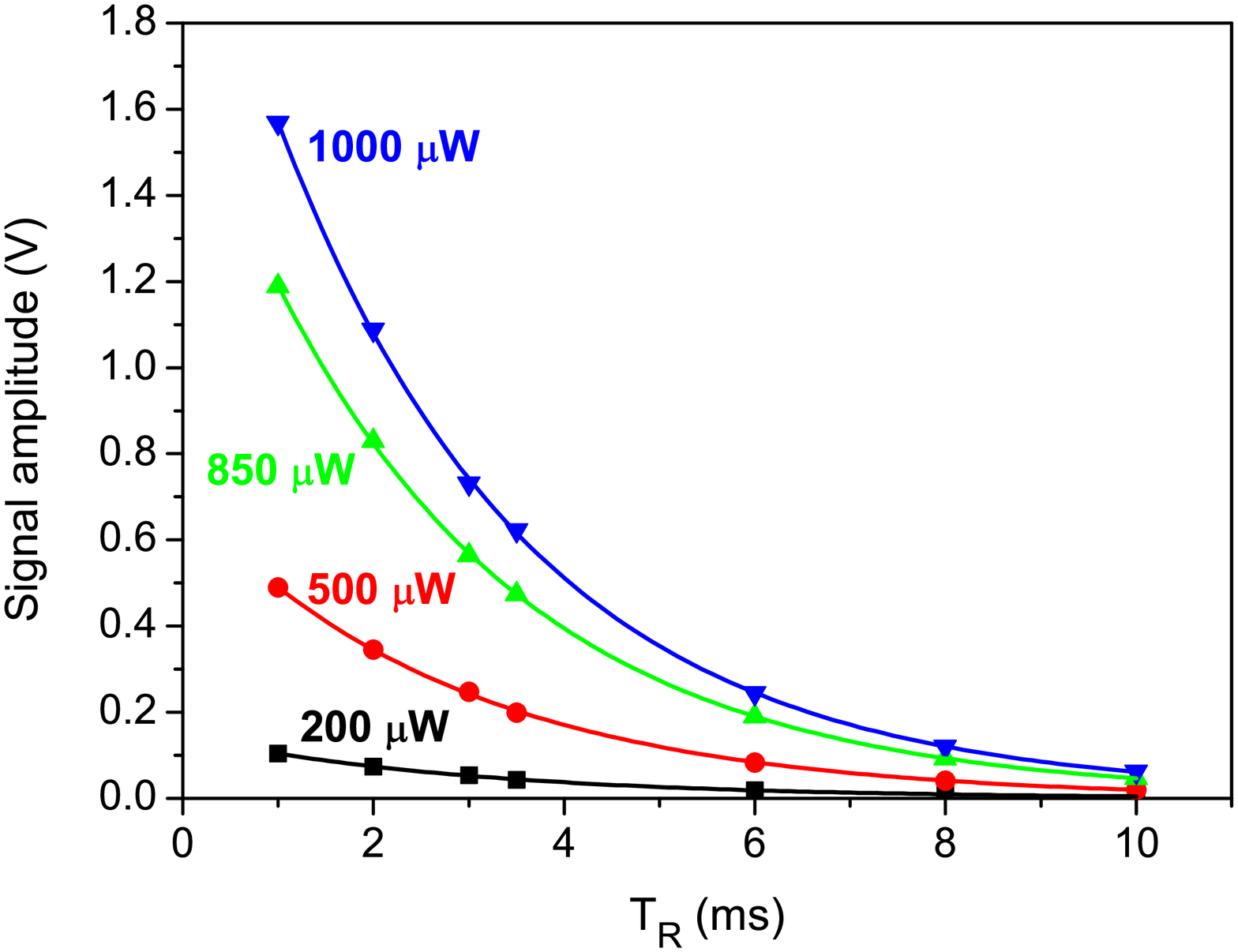}
\label{fig:signal-TR}} \vfill
\subfigure[]{\includegraphics[width=0.5\linewidth]{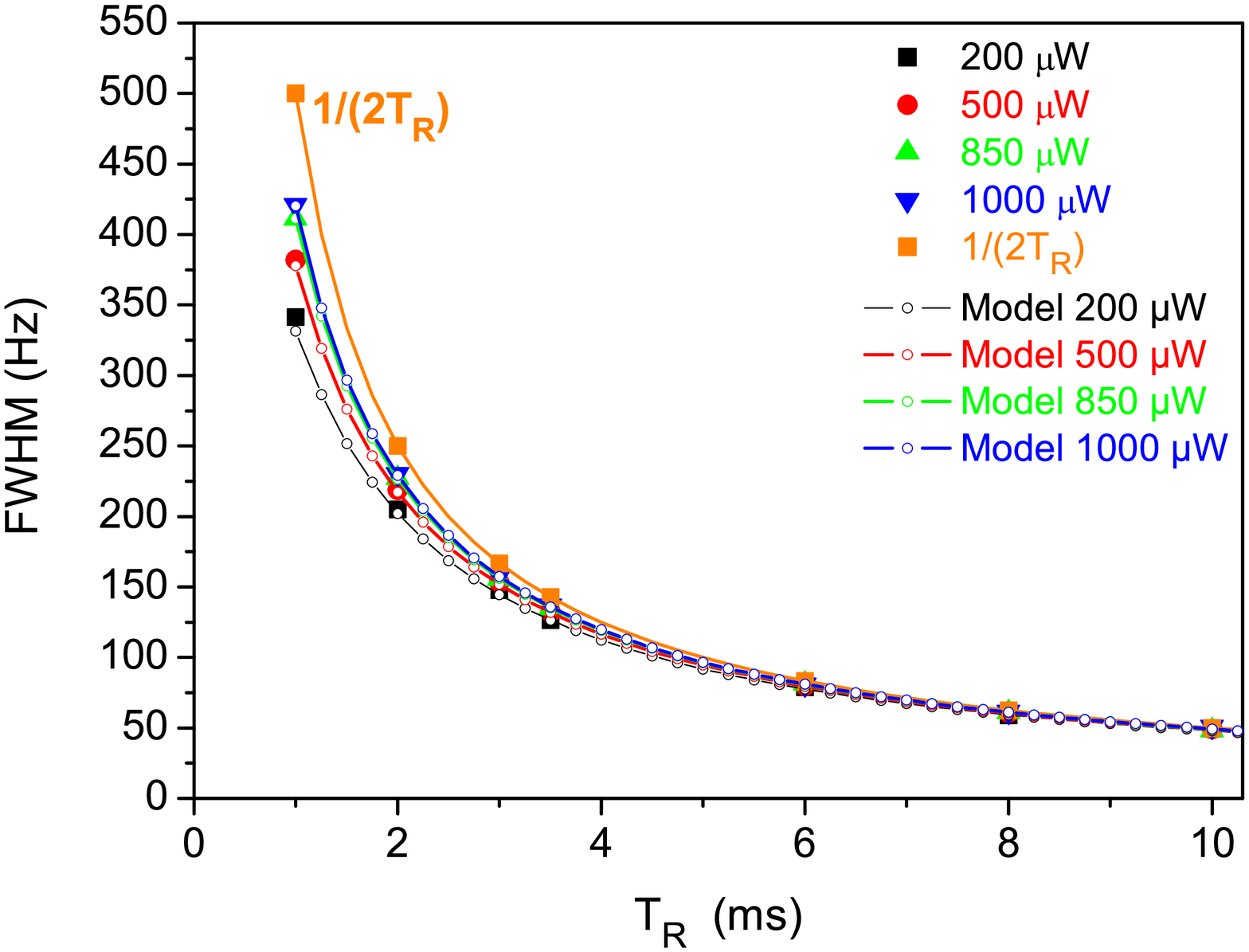}
\label{fig:fwhm-TR}} \vfill
\subfigure[]{\includegraphics[width=0.5\linewidth]{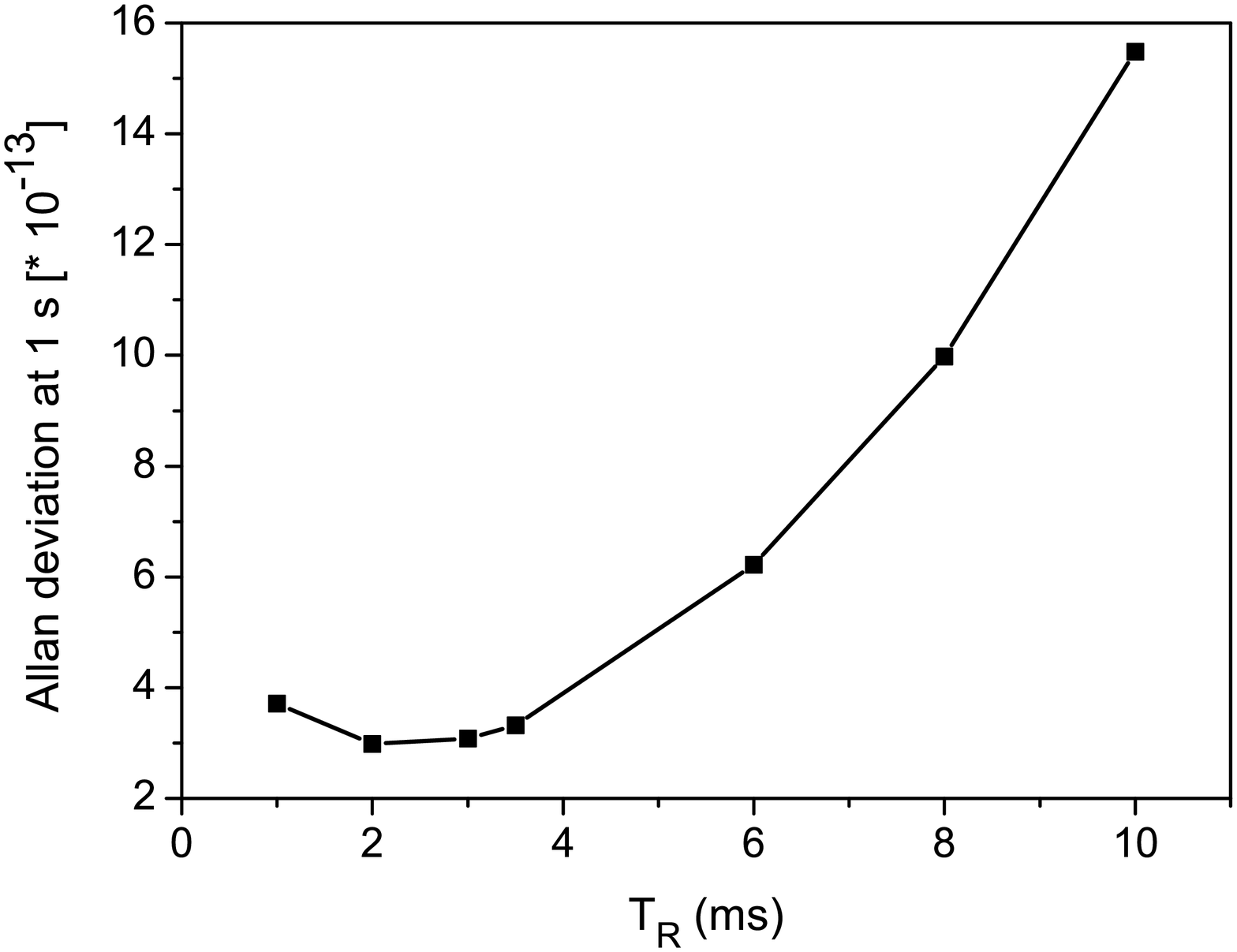}
\label{fig:stab-TR}}
 \caption{}
\end{figure*}

\begin{figure*}[t]
\centering
\subfigure[]{\includegraphics[width=0.5\linewidth]{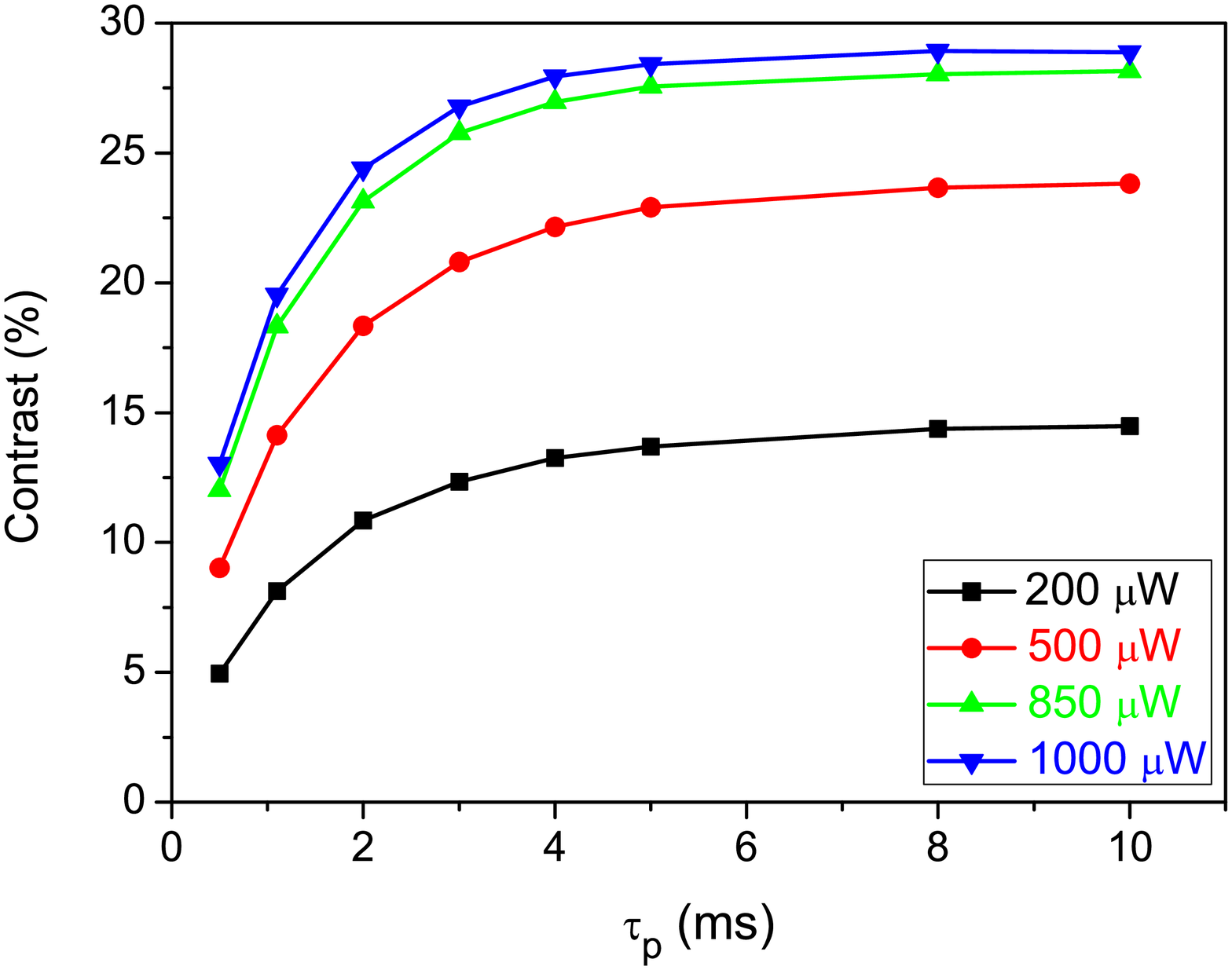}
\label{fig:C-taup}} \vfill
\subfigure[]{\includegraphics[width=0.5\linewidth]{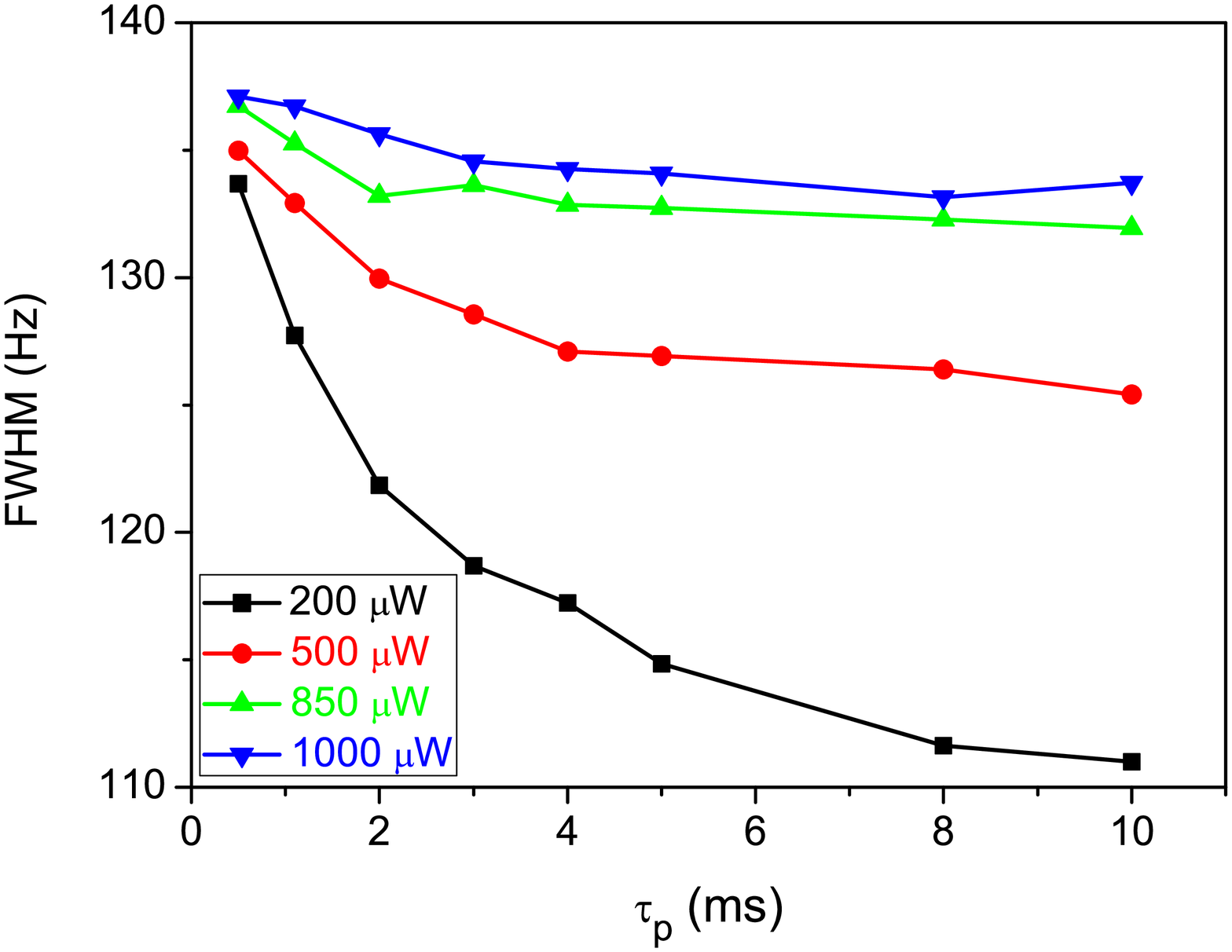}
\label{fig:fwhm-taup}} \vfill
\subfigure[]{\includegraphics[width=0.5\linewidth]{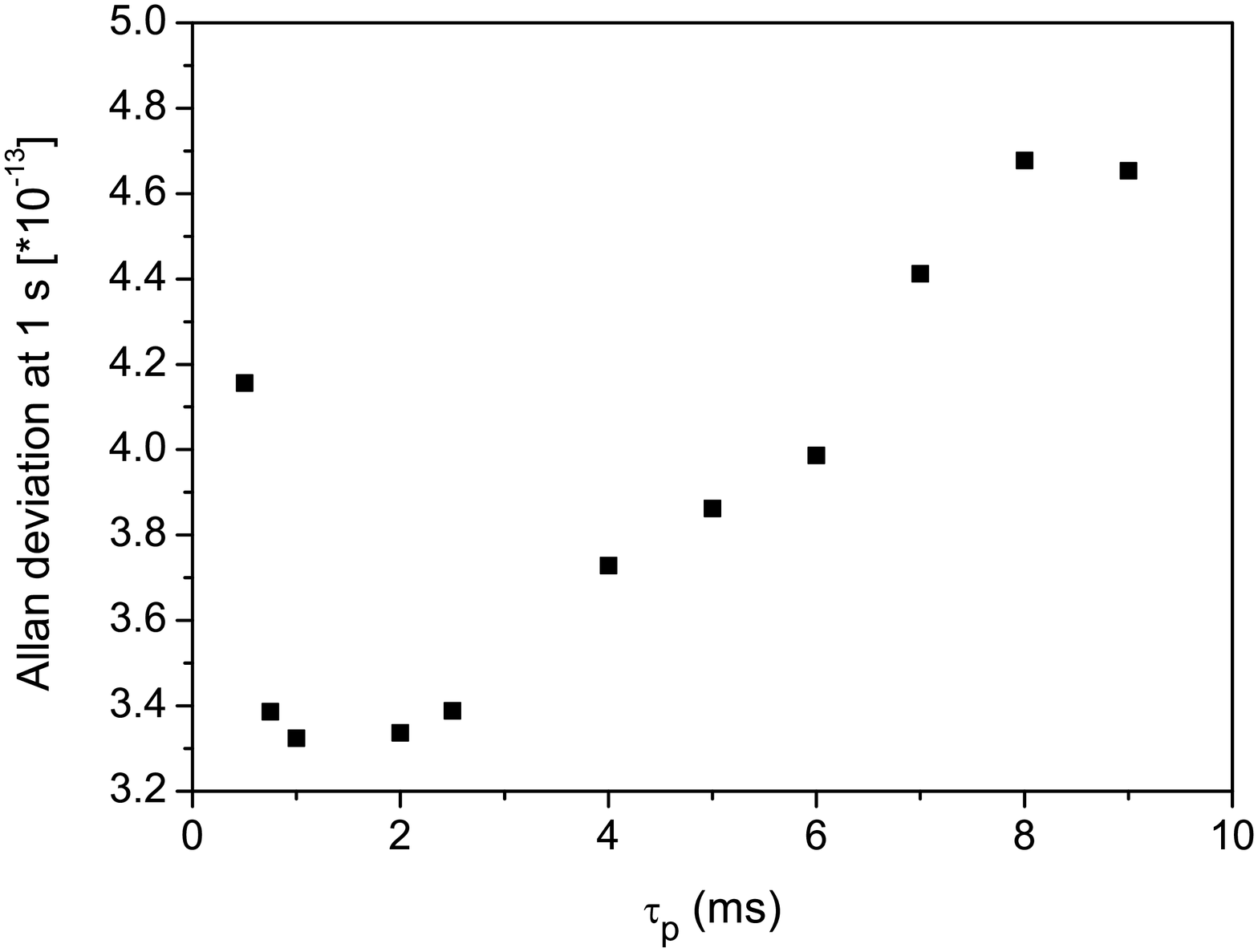}
\label{fig:stab-taup}}
 \caption{}
\end{figure*}

\begin{figure*}[t]
\centering
\includegraphics[width=\linewidth]{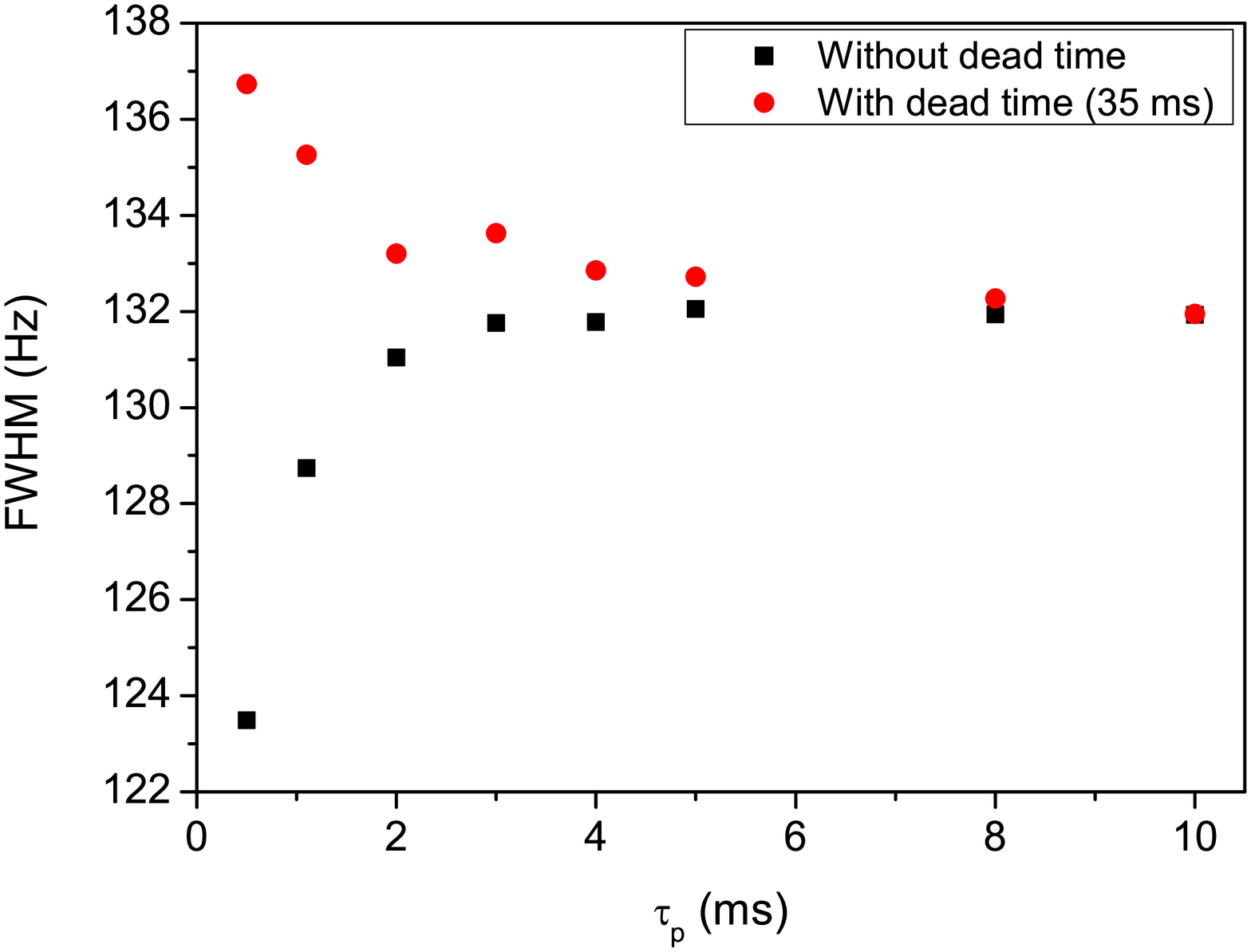}
\caption{}
\label{fig:impact-Tmort}
\end{figure*}

\begin{figure*}[t]
\centering
\includegraphics[width=\linewidth]{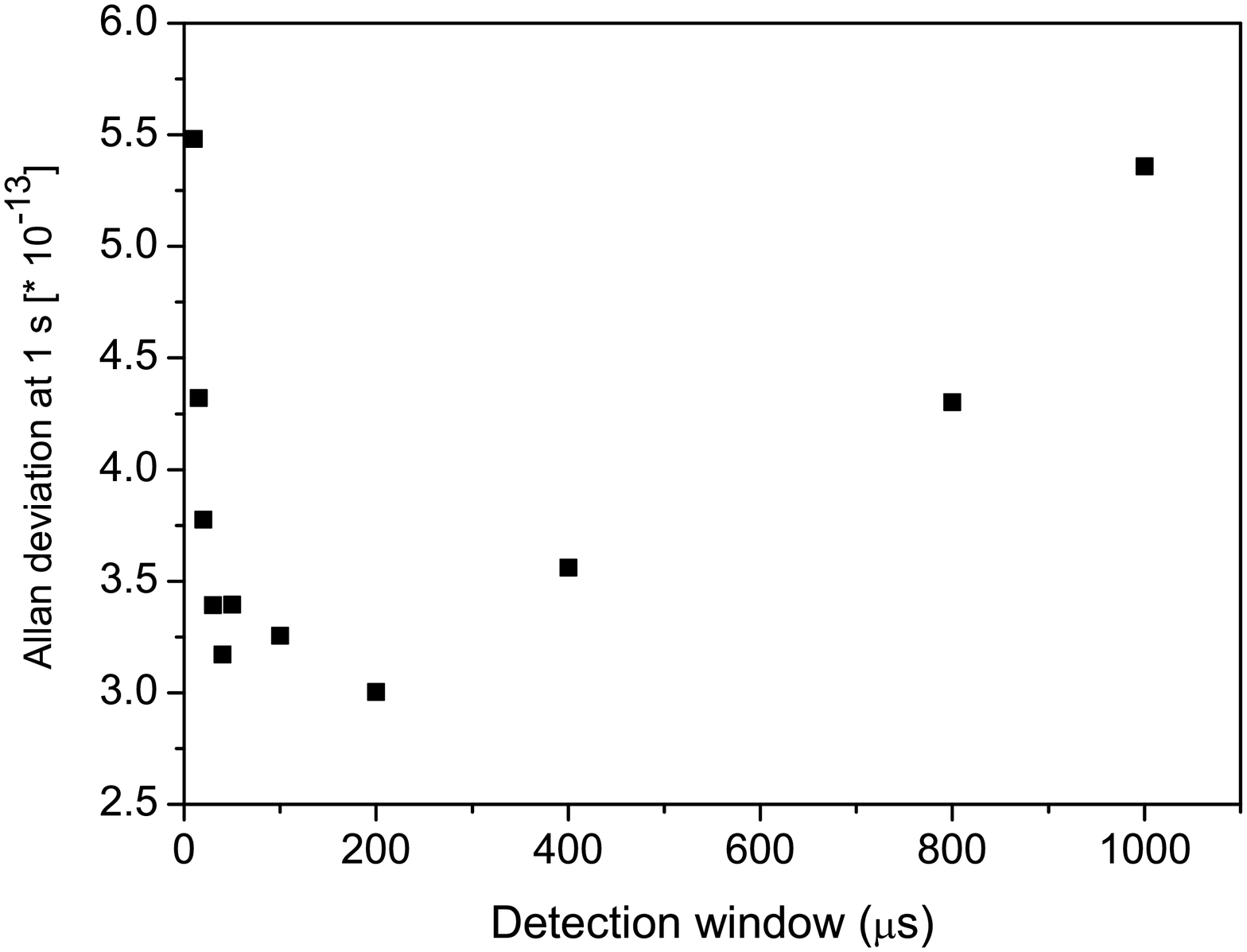}
\caption{}
\label{fig:stab-td}
\end{figure*}

Figure \ref{fig:fringe} shows typical Ramsey-CPT fringes, detected in the CPT cell for a total input laser power $P_i$ of 850 $\mu$W and a free-evolution time $T_R$ of 3.5 ms, with different span windows. The central fringe line-width is about 134 Hz. The CPT contrast, defined as the ratio between the central fringe peak-peak amplitude ($S$ = 0.48 V) and the dc level at half-height of the central fringe ($y_0$ = 2.85 V), is about 17 \%.\\
Figure \ref{fig:fwhm-cw-pulsed} shows the line-width of the central fringe versus the total laser power $P_i$ incident in the cell for different values of the free-evolution time $T_R$. For comparison, CPT line-width measurements obtained in the continuous-regime are reported. In the pulsed case, the central fringe line-width is mostly measured narrower than the expected Ramsey line-width given by $1/(2T_R)$, especially for low values of $T_R$ and at low laser power, and tends to $1/(2T_R)$ at high power. In the power range studied here, for $T_R =$ 1 ms, a non-negligible power broadening of the CPT Ramsey-fringe is observed, yielding 341 Hz at 200 $\mu$W and about 426 Hz at 1100 $\mu$W, i. e. an increase of 20 \% of the fringe line-width. This effect is less visible for higher values of $T_R$. This behaviour is not so surprising. In a two-level atom \cite{VA}, the $1/(2T_R)$ line-width of the central Ramsey fringe is known to be a valid approximation only if $T_R \gg \tau$, where $\tau$ is the length of each pulse. Otherwise, the width is narrower than $1/(2T_R)$, it depends on the pulse durations and on the Rabi frequency. In a three-level system, like in CPT, the same behaviour is expected. In this configuration, the resonance signal can be considered as the product of a broad Lorentzian resonance, whose width is driven by the first pulse length and the Rabi frequency, multiplied by a fast oscillating function of width $1/(2T_R)$. In the case where the $1/(2T_R)$ width can not be assumed to be small compared to the broad resonance line-width, the observed line-width of the central fringe is a complicated function of all the parameters \cite{Zanon:PRA:2011} but the effect of the broad resonance is to reduce the apparent fringe width. Since the broad resonance widens with the laser power, its effect on the fast oscillation is reduced with increased power and the observed fringe width then increases towards the $1/(2T_R)$ limit. Since we do not know exact analytical expression to describe this behaviour in the CPT case, we have performed a numerical simulation based on optical Bloch equations, applied to a three-level atom in presence of two resonant light fields. Parameters of the simulation are the Rabi frequency $\omega_R$ of both optical transitions (assumed equal if use of equal laser intensities on the Cs D$_1$ line, $\omega_R = 2.3 \times 10^6 \sqrt{P_i/S_c}$ in rad/s with $P_i$ the total laser power and $S_c$ the laser beam section area), the experimental Ramsey-CPT sequence parameters $\tau_p$, $T_R$, $\tau_D$, the spontaneous relaxation rate $\Gamma$ (such that $\Gamma/2\pi$ = 4.6 MHz corresponding to the excited state natural linewidth), the relaxation rate of the excited state $\Gamma^{\star}$ taking into account the buffer-gas induced optical broadening \cite{Pitz:PRA:2009} (such that $\Gamma^{\star}/2\pi$ = 333 MHz), the relaxation of optical coherences ($\gamma_o= \Gamma^{\star}/2$), and the relaxation rate of hyperfine coherence and ground-state populations ($\gamma = 1 / T_2$, with $T_2$ $\sim$ 3 ms). Calculation results of this very simplified model, reported in Fig. \ref{fig:fwhm-cw-pulsed} for several values of $T_R$, are in correct agreement with experimental data. Despite the power broadening in the pulsed regime (especially for low values of $T_R$), we note that the resonance line-width and the power broadening are significantly reduced in the pulsed regime compared to the continuous interrogation case. In the latter regime, as shown on Fig. \ref{fig:fwhm-cw-pulsed}, experimental data of the line-width (FWHM) $\Delta \nu$ are well-fitted by a linear function such as $\Delta \nu$ [Hz] $= 225 + 1.16~P_i$ with $P_i$ in $\mu$W.\\
Figure \ref{fig:c-p-TR} shows the contrast of the Ramsey-CPT central fringe versus the laser power for several values of $T_R$. The CPT contrast obtained in the continuous regime is also given for comparison. In the CW regime, the contrast is increased from 31 to about 53 \% in the studied laser power range. A saturation plateau seems to be reached for laser powers higher than 900 $\mu$W. In the pulsed case, the fringe contrast is reduced with increased values of $T_R$ because of relaxation of the hyperfine coherence. For $T_R =$ 3.5 ms, the fringe contrast is increased from 8 to 17 \% in the 200 - 1100 $\mu$W range.
We can assume that the fringe amplitude scales as the resonance amplitude obtained in the CW interrogation case. From \cite{Vanier:PRA:1998}, we deduce that the resonance amplitude $S$, for a three-level atom at steady state, is such that:
\begin{equation}
S \simeq \frac{\omega_R^4/\Gamma^{*2}}{\gamma+\omega_R^2/\Gamma^*},
\label{amplcw}
\end{equation}
$\omega_R^2$ is proportional to the laser power. As the background level scales as the laser power $P_i$, we can fit the experimental contrast data by $C=A P_i/(225+b P_i)$, where $A$ and $b$ are the free parameters and 225 is the relaxation term given by the fit of the CW linewidth versus the laser power. The agreement with the experimental data is satisfactory. From results obtained in Figs \ref{fig:fwhm-cw-pulsed} and \ref{fig:c-p-TR}, the authors note that the fringe contrast/FWHM ratio is increased with laser power by a factor 2.5 from 200 to 1100 $\mu$W for $T_R =$ 3.5 ms.\\
Figures \ref{fig:signal-TR} and \ref{fig:fwhm-TR} report the central fringe amplitude and line-width versus the free-evolution time $T_R$ for several values of the laser power. The pumping time is 1.1 ms and the detection window is 50 $\mu$s. On Fig. \ref{fig:signal-TR}, experimental data of the fringe amplitude are well-fitted by an exponential decay function, with a time constant of about 2.8 ms. The latter can be interpreted as an estimation of the CPT coherence lifetime $T_2$ in the cell. On Fig. \ref{fig:fwhm-TR}, we observe again for low values of $T_R$ that the central fringe line-width can be narrower than the expected line-width given by $1/(2T_R)$.
This effect is the same than the one explained previously. Our numerical simulations based on a density-matrix calculation for a three-level atom are in good agreement with experimental data. Figure \ref{fig:stab-TR} reports the clock short-term fractional frequency stability at 1 s averaging time versus the free-evolution time $T_R$. The clock frequency stability is optimized for a plateau with $2.8< T_R <3.5$ ms, i.e. for $T_R \sim T_2$. This behaviour has been already reported in numerous articles \cite{Guerandel:IM:2007, MAH:JAP:2015, Micalizio:Metrologia:2012}.\\
Figures \ref{fig:C-taup} and \ref{fig:fwhm-taup} show respectively the central fringe contrast and line-width versus the pumping time for several values of the laser power $P_i$. Experimental parameters are $T_R$ = 3.5 ms and $\tau_D$ = 50 $\mu$s. The contrast is increased with the pumping time and reaches a plateau for $\tau_p \sim$ 4 ms. Simultaneously, the fringe line-width is decreased when the pumping time increases to reach progressively a minimum value. This minimum value is reached for higher pumping time values when the laser power is reduced. Looking at Figs \ref{fig:C-taup} and \ref{fig:fwhm-taup}, we could expect that the clock Allan deviation is optimized for pumping times higher than 4-5 ms. However, as shown on Fig. \ref{fig:stab-taup}, we found that the clock short-term frequency stability is optimized for pumping times in the 1-2 ms range. For clarification, we report on Fig. \ref{fig:impact-Tmort} the evolution of the central fringe line-width with the pumping time in two different sequence parameters conditions. In the first case, during the Ramsey-CPT fringe spectroscopy, a dead-time of 35 ms ($\sim$ 10 - 12 $T_2$) is applied between each acquisition point. In the second case, the sequence is closer to the one used in clock locked configuration. The dead time is suppressed, the LO frequency is swept more rapidly and the transmitted power through the CPT cell is measured every clock cycle, i. e every 4.6 ms. For both tests, the Ramsey time $T_R$ is 3.5 ms and the laser power $P_i$ is 850 $\mu$W. With a slow scan of the LO frequency, the fringe line-width is reduced with increased pumping time for low values of the pumping time. The behaviour is totally opposite when lecture of the transmitted power is performed every clock cycle, without any dead-time between two acquisition points. For low values of $\tau_p$, it is worth to note that the steady-state is not reached at the end of the first pulse. Consequently, when the Ramsey fringes scan is fast and that no dead-time is applied between two acquisition points, the hyperfine coherence is not fully relaxed at the end of the free-evolution time and the new coherence builds up from the previous one. This is equivalent to a longer pumping time yielding a still narrower recorded fringe. For $\tau_p >$ 5 ms, both configurations give the same fringe line-width value because the steady-state is reached at the end of each pulse. Figure \ref{fig:stab-td} shows the impact of the detection window length on the clock short-term frequency stability. The clock Allan deviation is optimized for a detection time window ranging from 50 $\mu$s to about 200 $\mu$s.

\subsubsection{Cell temperature}
\begin{figure*}[t]
\centering
\includegraphics[width=\linewidth]{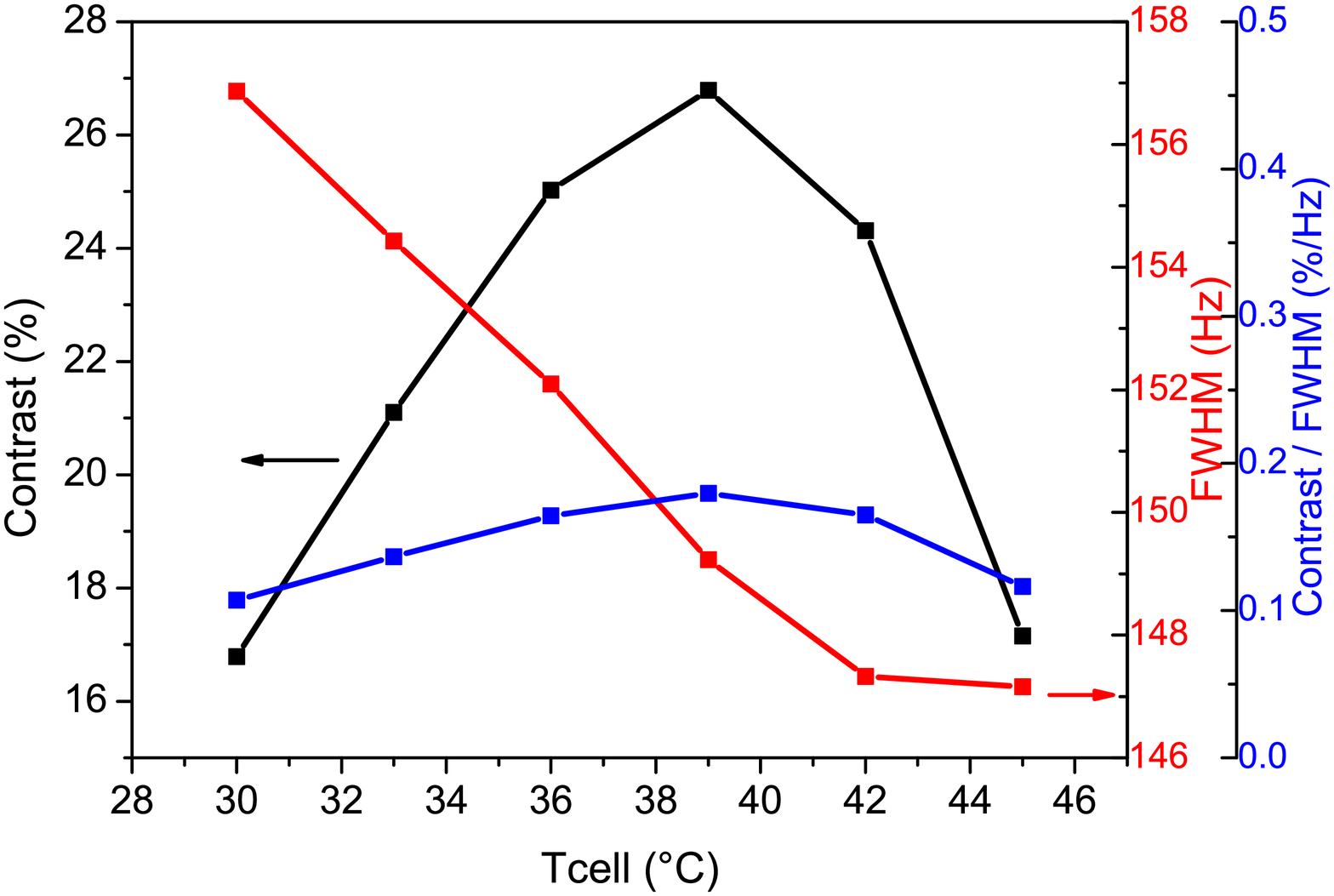}
\caption{}
\label{fig:fringe-Tcell}
\end{figure*}

\begin{figure*}[t]
\centering
\includegraphics[width=\linewidth]{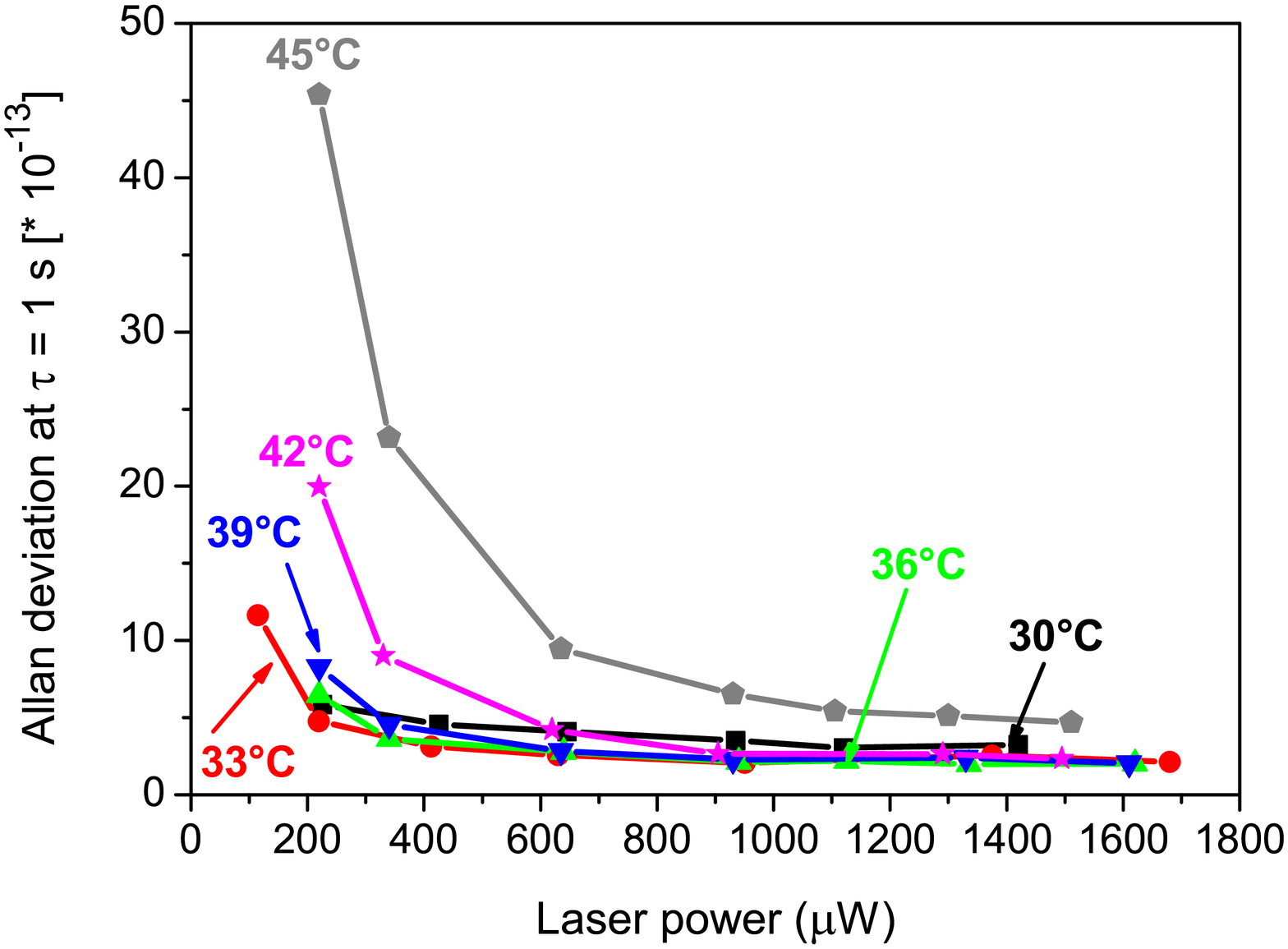}
\caption{}
\label{fig:stab-Pin-T}
\end{figure*}

Figure \ref{fig:fringe-Tcell} reports the central fringe contrast, line-width and contrast/linewidth ratio versus the cell temperature. The laser power is 863 $\mu$W while $T_R =$ 2.7 ms and $\tau_p =$ 1.1 ms. The fringe contrast is found to be maximized for a temperature of about 38$^{\circ}$C. The fringe line-width is slightly decreased with increased cell temperature. The ratio contrast/linewidth is optimized for a temperature of about 38$^{\circ}$C. Nevertheless, we note that the variation of this ratio is quite modest (less than a factor 2) in the studied temperature range.\\
We report on Fig. \ref{fig:stab-Pin-T} the clock Allan deviation at 1 s integration time for different laser power values at several cell temperatures. In the temperature range studied here, the short-term frequency stability is optimized in the pulsed case for laser powers higher than about 700 - 800 $\mu$W and reaches after a plateau where no further improvement is achieved. Similar optimal stabilities of about 2-3 $\times$ 10$^{-13}$ at 1 s averaging time are demonstrated in the 33 - 42$^{\circ}$C temperature range.

\subsubsection{Short-term frequency stability}
\begin{figure*}[t]
\centering
\includegraphics[width=\linewidth]{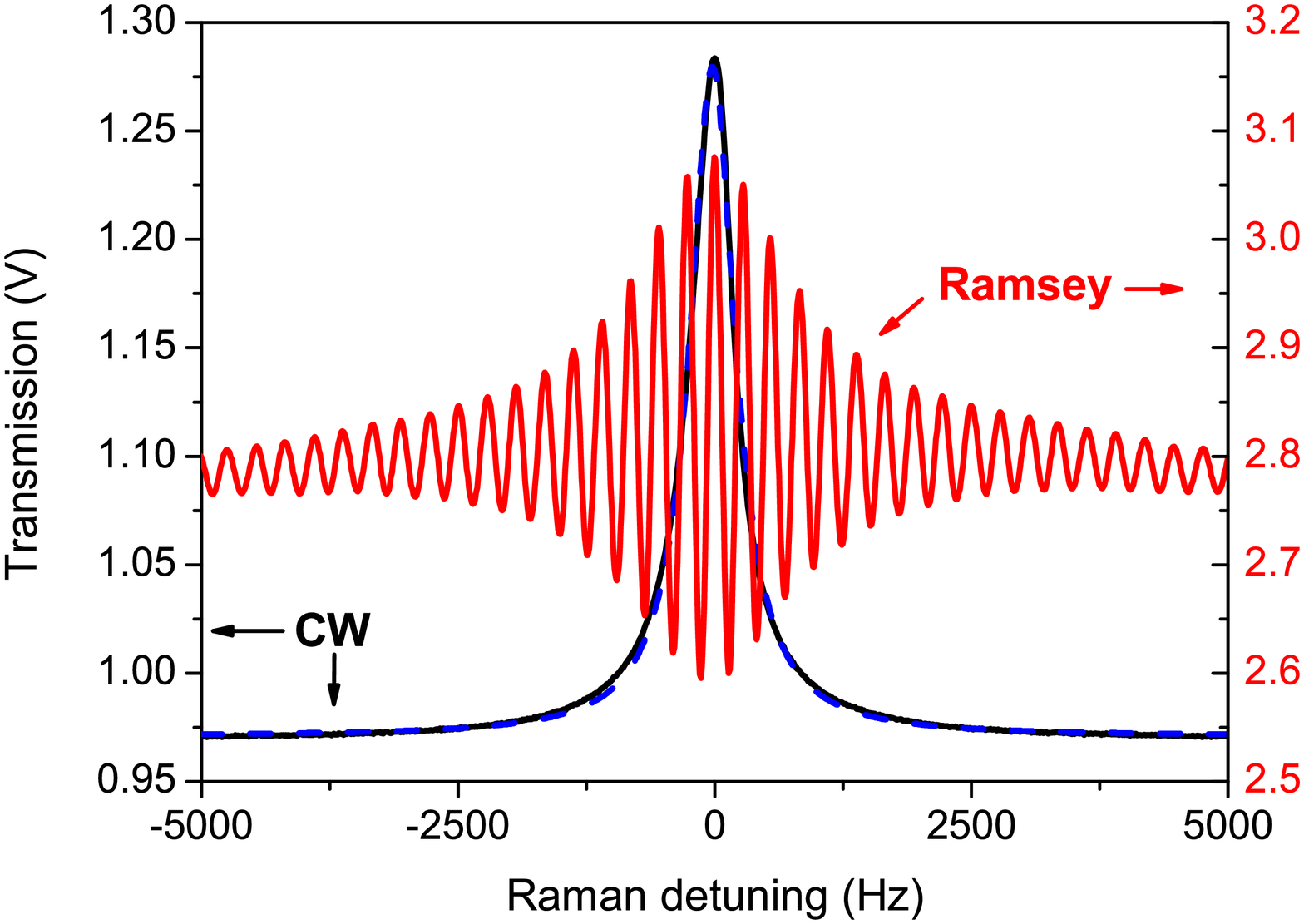}
\caption{}
\label{fig:cw-vs-pulsed}
\end{figure*}

\begin{figure*}[t]
\centering
\includegraphics[width=\linewidth]{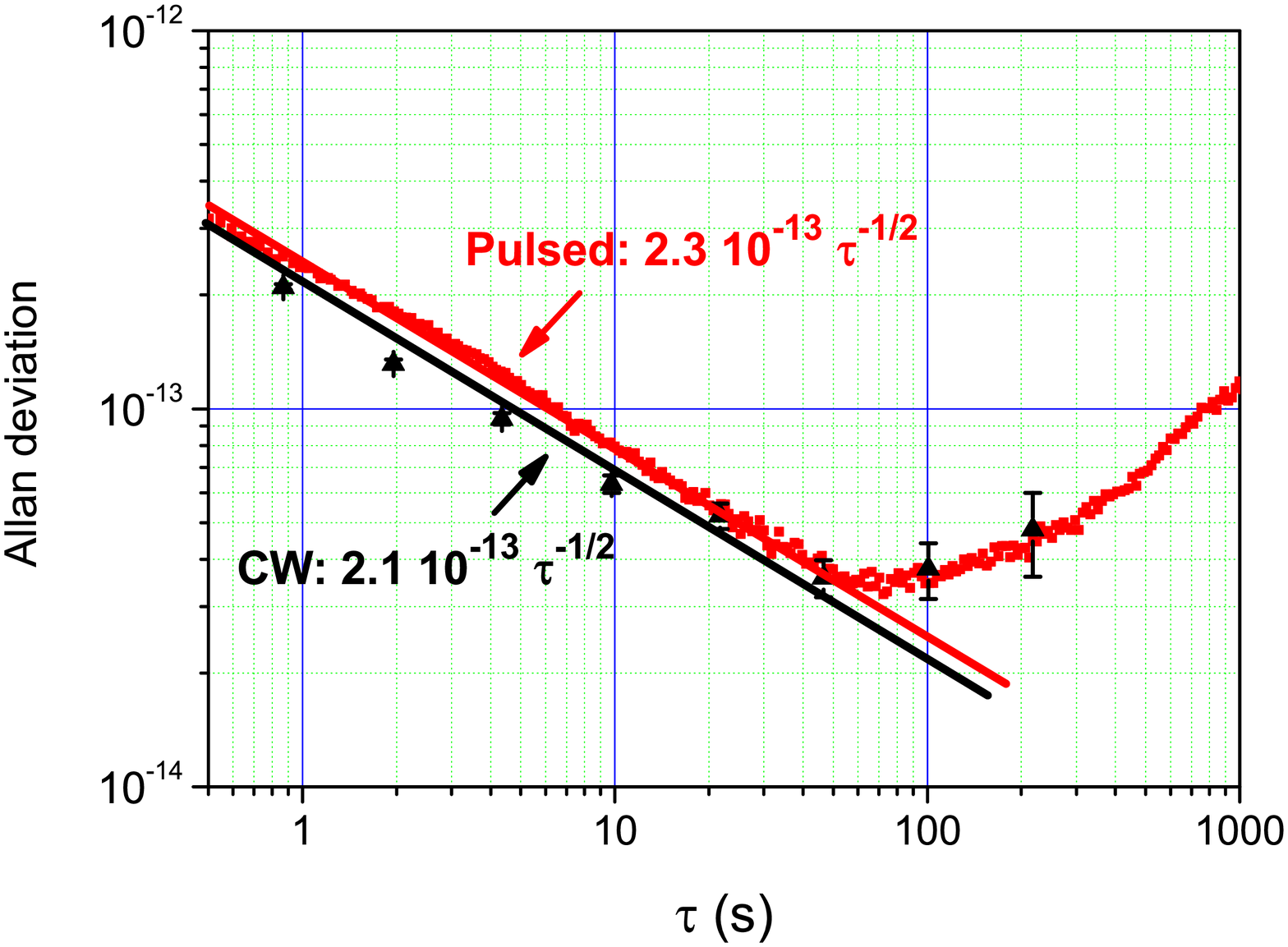}
\caption{}
\label{fig:allan}
\end{figure*}

Measurements of the clock short-term fractional frequency stability were performed for comparison in both CW and pulsed regimes, with the parameter values optimizing the stability in respective cases. In the CW regime, an AOM-based laser power stabilization system, as described in \cite{MAH:JAP:2015}, was implemented. In clock closed loop operation, the LO modulation frequency $f_M$ is 125 Hz, the modulation depth is $\pm$ 80 Hz, the cell input laser power $P_i$ is 270 $\mu$W and the cell temperature $T_{cell}$ is 35$^{\circ}$C. In the pulsed regime, no laser power stabilization is applied to date. Experimental parameters are $T_c$ = 4.6 ms, $T_R$ = 3.5 ms, $\tau_p$ = 1.1 ms, $\tau_D$ = 50 $\mu$s, $P_i$ = 850 $\mu$W and $T_{cell}$ = 35$^{\circ}$C.\\
Figure \ref{fig:cw-vs-pulsed} shows typical recorded clock signals in both CW and pulsed regimes. In the CW regime, the signal amplitude is $S$ = 0.31 V, the resonance line-width is $\Delta \nu$ $\sim$ 538 Hz, yielding a frequency discriminator slope roughly estimated by $S_l = S/\Delta \nu$ = 5.7 $\times$ 10$^{-4}$ V/Hz and the resonance contrast is $C$ = 31.8 \%. In the pulsed regime, we obtain a central fringe with an amplitude $S$ = 0.48 V, a line-width $\Delta \nu =$ 134 Hz, yielding $S_l$ = 3.6 $\times$ 10$^{-3}$ V/Hz and a contrast $C$ of 17 \%. These characteristics in both regimes are resumed in Table \ref{noise}.\\
Figure \ref{fig:allan} shows the Allan deviation of the clock frequency in both CW and pulsed regimes. The clock short-term fractional frequency stability is measured to be $2.1 \times 10^{-13} \tau^{-1/2}$ and $2.3 \times 10^{-13} \tau^{-1/2}$ for integration times up to a 100 seconds in the CW regime (with laser power servo) and pulsed case respectively. These performances are slightly better than those reported in \cite{MAH:JAP:2015} and close to those of best vapor cell atomic frequency standards \cite{Micalizio:Metrologia:2012, Kang:JAP:2015}. In the present study, comparable performances are obtained in the CW and the pulsed regime. Such conclusions were also reported in \cite{Gharavipour:FSM:2016, Novikova:JOSAB:2016}. \\
The short-term frequency stability of an atomic clock results from a trade-off between the resonance-based frequency discriminator slope $S_l$ and the detection noise. In the pulsed case, the sequence parameters have also to be taken into account. Table \ref{noise} resumes and compares main experimental conditions, clock resonance characteristics and noise contributions to the clock short-term fractional frequency stability, in both CW and pulsed regimes. The definition and estimation of noise sources contributions, supported by experimental measurements, were performed following the methodology described in \cite{Yun:PRAp:2017}. For sake of simplicity, we consider only white noise sources and we assume that the different noise contributions can independently add. The total expected Allan variance $\sigma_y^2 (\tau)$ is computed as the sum of $\sigma^2_{y,LO}$ (contribution due to the phase noise of the local oscillator) and $\sum_i \sigma^2_{y,p_i}$, with $\sigma^2_{y,p_i}$ the Allan variance of the clock frequency induced by the fluctuations of the parameter $p_i$ \cite{Yun:PRAp:2017}.\\
A correct agreement is found between the measurements and calculations. In the CW regime, the main limitation of the clock short-term frequency stability is the laser-induced amplitude modulation-to-frequency modulation (AM-FM) effect, caused by the sensitivity of the clock frequency to laser power variations. Following contributions are the laser amplitude modulation-to-amplitude modulation (AM-AM) noise (detection signal AM noise induced by the laser AM noise) and the LO phase noise \cite{Francois:RSI:2014}. We note that, without laser power stabilization, the laser AM-FM noise contribution is increased by a factor 2-3 while the AM-AM noise contribution is increased at the level of 1.9 $\times$ 10$^{-13}$. In this case, the clock short-term fractional frequency stability was found to be degraded by a factor of 2-3, yielding 4-6 $\times$ 10$^{-13}$ at 1 s. In the pulsed case, the main contribution to the noise budget is to date the laser AM-AM noise process at the level of 2.7 $\times$ 10$^{-13}$, followed by the laser FM-AM conversion process (detection signal AM noise induced by the laser carrier frequency noise), the LO phase noise and the laser AM-FM process. Other contributions are well below and negligible at the moment. With laser power stabilization, assuming a gain on the detection noise similar to the one obtained in the CW regime, the laser AM-AM contribution could be reduced in the pulsed case at the level of 4.8 $\times$ 10$^{-14}$ instead of 2.7 $\times$ 10$^{-13}$ presently. This would lead in the pulsed case to an overall fractional frequency stability at the level of 1 $\times$ 10$^{-13}$ at 1 s averaging time.\\
In the present configuration, pulsed and continuous regimes show similar short-term frequency stability levels. Indeed, according to Table I, we notice that the contribution to the clock frequency stability of the laser AM-FM noise is 4.4 times lower in the pulsed case than in the CW case, due to a larger light shift coefficient in the latter case. This gain in the pulsed regime is counter-balanced by the laser AM-AM noise contribution, 4.5 times higher than in the CW regime. This is due to a large extent to a higher AM-noise level in the pulsed regime (a factor of 28 with respect to the CW regime), because of the absence of power stabilization and a higher incident laser power into the cell, counter-balanced by a larger frequency discriminator (6.3 times higher in the pulsed regime)

\subsection{Preliminar investigations on light-shift effects}\label{sec:sts}
\begin{figure*}[t]
\centering
\subfigure[]{\includegraphics[width=0.8\linewidth]{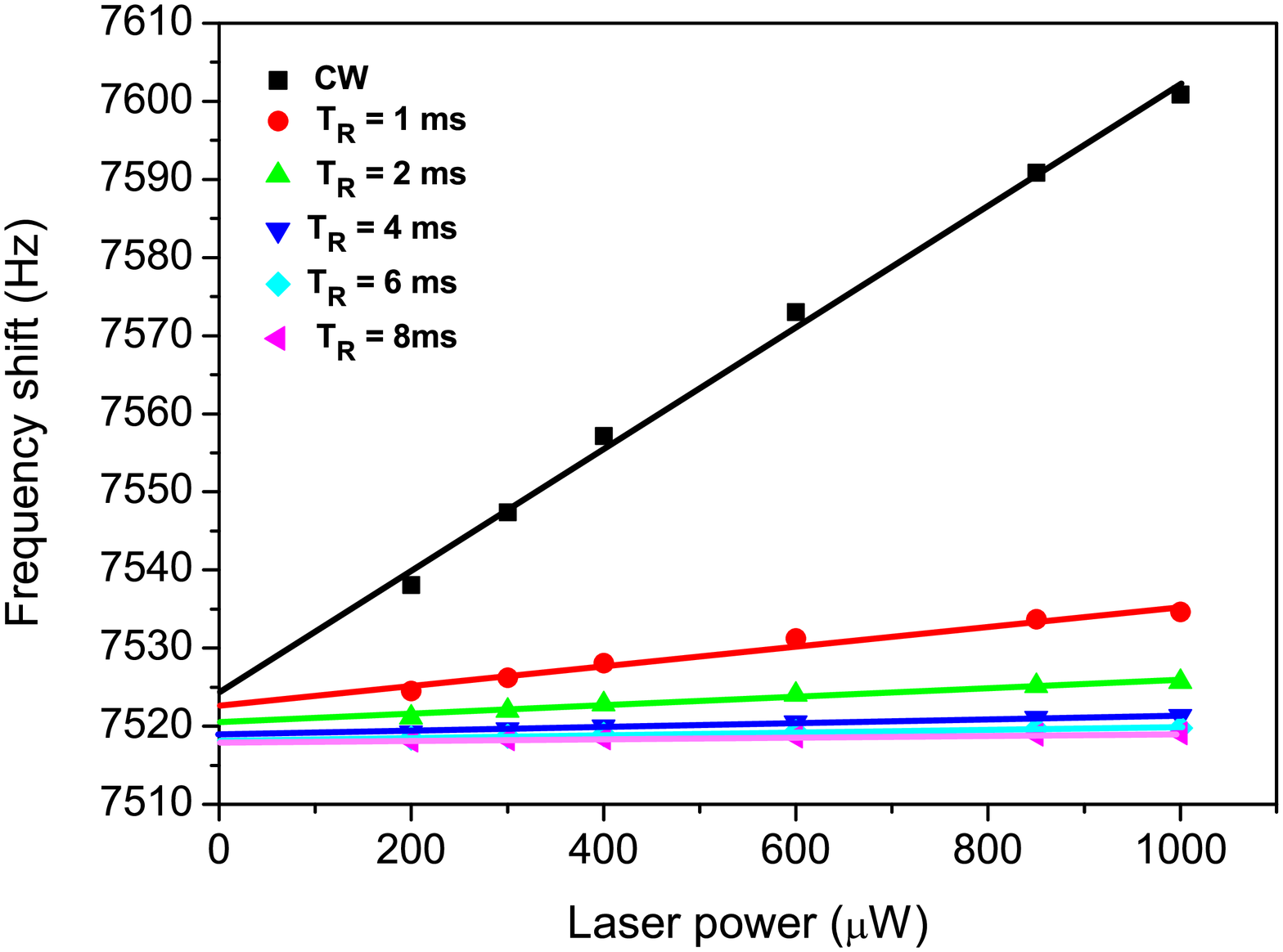}
\label{fig:ls-TR}} \vfill
\subfigure[]{\includegraphics[width=0.8\linewidth]{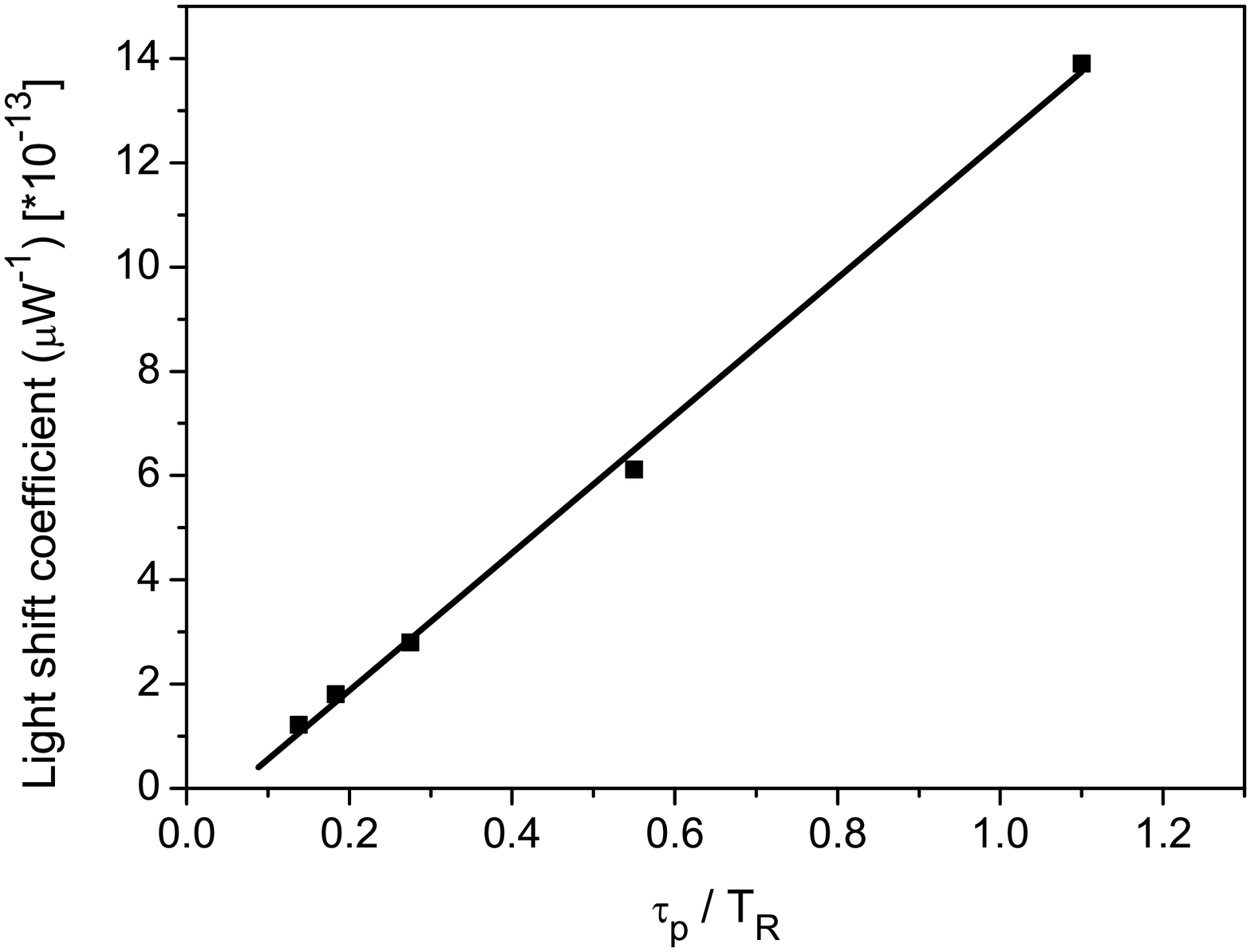}
\label{fig:lsslope-TR}}
\caption{}
\end{figure*}

A major and common issue in most vapor cell clocks is the degradation of their fractional frequency stability for integration times typically higher than 100 or 1000 s. This characteristic is a serious obstacle to their deployment in practical applications. This degradation is generally caused by laser intensity and frequency light-shift effects, cell temperature or pressure effects. In general, light shift is recognized as a major cause and is being studied with significant interest \cite{Yudin:PRA:2010, Miletic:APB:2012, Huntemann:PRL:2012, Kozlova:IM:2014, Yano:PRA:2014, Pati:JOSAB:2015, Blanshan:PRA:2015, Hobson:PRA:2016, Zanon:Arxiv:2016, Yudin:Arxiv} for improvement of atomic frequency standards. As reported in previous literature \cite{Castagna:UFFC:2009, Boudot:IM:2009}, the pulsed interaction presents the advantage to reduce significantly the sensitivity of the clock frequency to laser intensity variations.\\
For this purpose, we decided to report a preliminary investigation of light-shift effects in this clock in the pulsed regime.
Figure \ref{fig:ls-TR} plots the clock frequency (frequency shift from the unperturbed Cs atom frequency 9.192 631 770 GHz) versus the laser power for several values of $T_R$. Results obtained in the continuous regime are reported for comparison. Experimental parameters are: $T_{cell}$ = 35$^{\circ}$C, $\tau_p$ = 1.1 ms and $\tau_D$ = 50 $\mu$s. In the CW regime, the intensity light-shift coefficient is measured to be 8 $\times$ 10$^{-12}$ $\mu$W$^{-1}$ (relative to the clock frequency). In the pulsed case, we observe a significant reduction of the light shift slope with increased Ramsey time, resumed on Fig. \ref{fig:lsslope-TR}. The light-shift slope is reduced to 6 $\times$ 10$^{-13}$ $\mu$W$^{-1}$ and 3 $\times$ 10$^{-13}$ $\mu$W$^{-1}$ for $T_R =$ 2 and 4 ms respectively. This behavior is qualitatively explained by the fact that with increased $T_R$, atoms spend a longer fraction of the clock cycle in the dark where they experience no or a greatly reduced light shift effect. In Ref. \cite{Borde}, C. Bord\'e reports for a 2-photon transition scheme that the light shift slope in a pulsed Ramsey scheme should be reduced compared to the continuous regime case by a factor proportional to $d/w_0$ where $d$ is the distance between both Ramsey cavity zones and $w_0$ the interaction length. In the present pulsed CPT clock experiment, the ratio $d/w_0$ is equivalent to the $T_R/\tau_p$ ratio. In our experimental case, with $\tau_p =$ 1.1 ms and $T_R =$ 3.5 ms, we observe that the light-shift slope is reduced by a factor of about 26 in the pulsed regime, much greater than the observed factor 3.5/1.1 = 3.2. This disagreement, related to the specificity of the Ramsey-CPT interrogation, will be studied in a near future.\\
In the present experiment, the sensitivity of the clock frequency to laser power variations is (in fractional value) at the level of 8 $\times$ 10$^{-12}$$\mu$W$^{-1}$ in the CW regime and reduced by a factor 26 to 3 $\times$ 10$^{-13}$$\mu$W$^{-1}$ in the pulsed case ($T_R$ = 3.5 ms). Simultaneously, the typical laser power used in optimal clock operation is a factor 3.14 lower in the CW case ($P_i$ = 270 $\mu$W) than in the pulsed case ($P_i$ = 850 $\mu$W). In that sense, we can expect that the pulsed interaction, compared to the CW case, should relax the constraints on the required relative laser power fluctuations by a factor of about $26/3.14 = 8.3$ (about one order of magnitude) to reach a given clock frequency stability level.\\
An ideal clock with a fractional frequency stability of $2 \times 10^{-13}~\tau^{-1/2}$ is expected to reach the 10$^{-14}$ level after about 400 s integration time. In the CW regime, from data given just above, this requires to demonstrate that laser power fluctuations $\sigma P_i$ at 400 s averaging time are lower than 1.25 nW, i.e. $\sigma P_i / P_i = 4.6 \times 10^{-6}$. In the pulsed case, this laser power control level would be reduced such that $\sigma P_i / P_i = 3.8 \times 10^{-5}$, which seems easier to achieve.\\
In that sense, the pulsed interaction is expected to be the best solution for the demonstration of a clock with improved mid-term and long-term frequency stability. Note that the pulsed interaction involves a bit more complex digital electronics design for proper operation of all required servo loops of the clock and the addition of a fast optical switch to generate the light pulsed sequence. However, these functions remain accessible without any significant increase of the consumption and cost of the device. Detailed investigations will be pursued in the future in the laboratory to improve the mid-term and long-term frequency stability performances of the present clock.

\section{Conclusions}\label{sec:conclu}
We have demonstrated a pulsed CPT-based Cs vapor cell atomic clock exhibiting a short-term fractional frequency stability of 2.3 $\times$ 10$^{-13} \tau^{-1/2}$ up to 100 seconds averaging time. These performances are comparable to those of best vapor cell microwave frequency standards developed worldwide. The clock short-term frequency stability was optimized through the adjustment of key experimental parameters including the cell temperature, the laser power and the Ramsey-CPT sequence parameters. A dependence on laser power of the central fringe in the pulsed regime was reported. The central fringe line-width was measured commonly narrower than the expected Ramsey line-width $1/(2T_R)$ for short free-evolution times. Similar frequency stability performances were obtained in the CW regime, with the use of an additional laser power stabilization system. A significant reduction of the clock frequency sensitivity to laser power variations was measured in the pulsed case, especially for high values of the free-evolution time $T_R$. We think that the use of a pulsed Ramsey-like interrogation, possibly combined with some extensions of Hyper-Ramsey spectroscopy and original synthetic frequency protocols \cite{Yudin:PRA:2010, Zanon:Arxiv:2016, Yudin:Arxiv} applied to CPT, improved laser power stabilization or advanced frequency drift compensation techniques, are exciting and promising methods to improve the long-term frequency stability performances of such CPT-based vapor cell clocks.

\section*{Acknowledgments}\label{sec:acknow}
This work has been funded by the EMRP program (IND55 Mclocks). The EMRP is jointly funded by the EMRP participating countries within EURAMET and the European Union. This work was partly supported by LNE, R\'egion de Franche-Comt\'e and LabeX FIRST-TF. The authors would like to thank C. Rocher and P. Abb\'e (FEMTO-ST) for help with electronics, P. Bonnay (Observatoire de Paris) for filling of the vapor cell and V. Maurice (FEMTO-ST) for significant contribution to the development of the clock control Python software program.

\clearpage
\section*{Tables}
\begin{table*}[h]
\begin{center}
\caption{\label{noise} \textnormal{Main characteristics of the clock resonance and contributions to the clock short term frequency stability at $\tau$ = 1 s. The noise sources contributions are named and calculated as described in \cite{Yun:PRAp:2017}. The local oscillator (LO) phase noise item describes the contribution of the LO phase noise to the clock Allan deviation through the intermodulation (CW case) or Dick effect (pulsed case) \cite{Francois:RSI:2014}. The laser AM-AM noise is the amplitude noise induced by the laser intensity noise. The laser FM-AM noise is the amplitude noise induced by the laser carrier frequency noise. The $P_{\mu w}$ item stands for contributions of the microwave power fluctuations to the clock Allan deviation. Laser AM-FM and FM-FM are laser-induced frequency-shift effects from laser power or laser frequency variations respectively. Other contributions, from the cell temperature (item $T_{cell}$) and magnetic field (item $B$), are much lower. In the CW regime, the laser power is stabilized whereas it is not in the pulsed case. }}
\begin{tabular}{|c|c|c|}
\hline
Regime & CW & Pulsed\\
\hline
$S$ (V) & 0.31 & 0.48\\
$\Delta \nu$ (Hz) & 538 & 134 \\
$S_l= S / \Delta \nu$ (V/Hz)& 5.7 $\times$ 10$^{-4}$ & 3.6 $\times$ 10$^{-3}$ \\
$C$ (\%)& 31.8 & 17 \\
$C S_l$& 1.8 $\times$ 10$^{-4}$ & 6.1 $\times$ 10$^{-4}$ \\
$P_i$ ($\mu$W) & 270 & 850 \\
$P_o$ ($\mu$W) & 56 & 205 \\
$T_{cell}$ ($^{\circ}$C) & 35 & 35 \\
$f_M$ or $f_c$ & 125 & 217 \\
\hline
\hline
Noise Source & $\sigma$~(1 s) $\times$ 10$^{13}$& $\sigma$~(1 s) $\times$ 10$^{13}$\\
\hline
Shot noise & 0.16 & 0.27\\
Detector noise & 0.11 &  0.10\\
LO phase noise & 0.5 & 0.6\\
Laser AM-AM & 0.6 & 2.7\\
Laser FM-AM & 0.39 & 0.49\\
$P_{\mu w}$ & 5 $\times$ 10$^{-2}$   & 5.7 $\times$ 10$^{-2}$\\
Laser AM-FM & 2.16&0.49 \\
Laser FM-FM & 2.7 $\times$ 10$^{-2}$&2.9 $\times$ 10$^{-2}$ \\
$T_{cell}$ & 8 $\times$ 10$^{-3}$ & 2.4 $\times$ 10$^{-3}$\\
$B$ & 3.8 $\times$ 10$^{-3}$ & 3.8 $\times$ 10$^{-3}$\\
\hline
Total (expected) & 2.4 & 2.9\\
Measured  & 2.1 & 2.3\\
\hline
\hline
\end{tabular}
\end{center}
\end{table*}

\clearpage
\section*{Figure captions}
\begin{enumerate}
\item\label{fig:1} (Color online) Schematic of the pulsed Cs vapor cell atomic clock based on push-pull optical pumping. DFB: Distributed feedback diode laser, FC: fiber collimator, MZ EOM: Mach-Zehnder electro-optic modulator, AOM: acousto-optic modulator, LO: microwave local oscillator, FPD: fast photodiode, PD1: photodiode, RF: radiofrequency source, bias: reference voltage, Michelson: Michelson-type delay-line and polarization orthogonalizer system, M: mirror, QWP: quarter-wave plate, HWP: half-wave plate, PC-DAQ: personal computer - data acquisition card. The inset shows the CPT diagram involved in the push-pull optical pumping technique.
\item\label{fig:2} (Color online) Typical sequence used to perform Raman-Ramsey spectroscopy. (a) spectroscopy (b) clock operation.
\item\label{fig:3} (Color online) Typical Ramsey-CPT fringes. Experimental parameters are $T_R$ = 3.5 ms, $\tau_D$ = 50 $\mu$s, $\tau_p$ = 1.1 ms, $P_i$ = 850 $\mu$W, $T_{cell}$ = 35$^{\circ}$C. The inset shows a zoom on Ramsey-CPT fringes on a total span of 1.5 kHz.
\item\label{fig:4} (Color online) (a) Central fringe line-width versus the laser power $P_i$ incident in the cell for several values of $T_R$ (1, 2, 3.5, 5 and 10 ms). Experimental parameters are : $T_{cell}$ = 35$^{\circ}$C, $\tau_p =$ 1.1 ms, $\tau_D =$ 50 $\mu$s. Experimental data are compared with a numerical model based on optical Bloch equations (see the article text). For information, the grey dashed and the pink dotted lines show the expected 1/(2$T_R$) line-width for $T_R =$ 1 or 3.5 ms respectively. Results obtained in the pulsed regime are compared with the CPT line-width measured in the CW regime. Experimental data in the CW case are fitted by a linear function. (b) Central fringe contrast versus the laser power $P_i$ incident in the cell for same values of $T_R$. Results in the CW regime are reported. Symbols: experimental data. Solid lines: fit (see the text).
\item\label{fig:5} (Color online) Fringe amplitude (a), FWHM (b) versus the free-evolution time $T_R$ for several values of the laser power $P_i$ (200, 500, 850, 1000 $\mu$W). Other parameters are $T_{cell}$ = 35$^{\circ}$C, $\tau_p$ = 1.1 ms and $\tau_D$ = 50 $\mu$s. For (a), experimental data are fitted by an exponential decay function. For (b), experimental data are compared to the $1/(2T_R)$ line-width. Experimental data are in correct agreement with a 3-level atom model calculation (see the article text). (c): Clock short-term fractional frequency stability at 1 s integration time versus $T_R$. The laser power $P_i$ is 850 $\mu$W.
\item\label{fig:6} (Color online) Central fringe contrast (a), FWHM (b) versus the pumping time $\tau_p$ for several values of the laser power $P_i$ (200, 500, 850, 1000 $\mu$W). Other parameters are $T_{cell}$ = 35$^{\circ}$C, $T_R$ = 3.5 ms and $\tau_D$ = 50 $\mu$s. A dead-time of 35 ms is applied between each acquisition point for (a) and (b). (c): Clock short-term frequency stability at 1 s integration time versus $\tau_p$. The laser power $P_i$ is 850 $\mu$W.
\item\label{fig:7} (Color online) Central fringe line-width versus the pumping time $\tau_p$ in two different conditions. In the first case, a dead-time of 35 ms is applied between each acquisition point. In the second case, the transmitted power through the cell is recorded every "clock" cycle, i.e every 4.6 ms.  Experimental parameters are $T_{cell}$ = 35$^{\circ}$C, $P_i$ = 850 $\mu$W, $T_R$ = 3.5 ms.
\item\label{fig:8} Clock Allan deviation at 1 s averaging time versus the detection window length. Experimental parameters are $T_R=$3.5 ms, $\tau_p=$1.1 ms, $T_{cell}$ = 35$^{\circ}$C, $P_i$ = 850 $\mu$W.
\item\label{fig:9} (Color online) Central fringe contrast, line-width and contrast/linewidth ratio versus the cell temperature. Experimental parameters are: $P_i$ = 863 $\mu$W, $T_R =$ 2.7 ms, $\tau_p =$ 1.1 ms, $\tau_D =$ 50 $\mu$s.
\item\label{fig:10} (Color online) Clock Allan deviation at 1 s integration time versus the laser power $P_i$ for several cell temperature values. Experimental parameters are: $T_R =$ 2.7 ms, $\tau_p =$ 1.1 ms, $\tau_D =$ 50 $\mu$s.
\item\label{fig:11} (Color online) Typical clock signals in CW and pulsed regimes. In the CW regime, the laser power is $P_i$ = 270 $\mu$W. The experimental spectrum is fitted by a Lorentzian function (in blue). In the pulsed regime, parameters are: $P_i$ = 850 $\mu$W, $T_R =$ 3.5 ms, $\tau_p =$ 1.1 ms, $\tau_D =$ 50 $\mu$s.
\item\label{fig:12} (Color online) Allan deviation of the clock frequency in both CW (black triangles) and pulsed (red squares) regimes. In the CW regime, the laser power is actively stabilized at $P_i$ = 270 $\mu$W, the local oscillator modulation frequency $f_M$ is 125 Hz and the LO modulation depth is $\pm$80 Hz. The dashed line is a fit to the curve with a 2.1 $\times$ 10$^{-13}$ $\tau^{-1/2}$ slope. In the pulsed case, experimental parameters are: $P_i$ = 850 $\mu$W, $T_R =$ 3.5 ms, $\tau_p =$ 1.1 ms, $\tau_D =$ 50 $\mu$s. The dashed line is a fit to the curve with a 2.3 $\times$ 10$^{-13}$ $\tau^{-1/2}$ slope.
\item\label{fig:13} (Color online) (a) Clock frequency shift (from the unperturbed Cs atom frequency) versus the laser power $P_i$ in the CW and pulsed regime for several values of $T_R$ (1, 2, 4, 6 and 8 ms). Experimental parameters are:  $T_{cell}$ = 35$^{\circ}$C, $\tau_p =$ 1.1 ms and $\tau_D =$ 50 $\mu$s. (b): Light-shift slope versus the $\tau_p / T_R$ ratio. Experimental data are fitted by a linear function.


\end{enumerate}

\clearpage

\clearpage

\clearpage


\clearpage




\clearpage

\clearpage

\clearpage

\clearpage

\clearpage

\clearpage


\clearpage

\clearpage

\clearpage

\clearpage




\begin{thebibliography}{1}

\bibitem{GPS}
R. T. Dupuis, T. Lynch and J. R. Vacaro, Proceedings of the 2008 IEEE International
Frequency Control Symposium edited by Jadusliwer B. (Honolulu, Hawaii-USA) 2008,
pp. 655-660.

\bibitem{Galileo}
P. Waller et al., IEEE Trans. Ultrason. Ferroelec. Freq. Control \bf{57}, \rm 3, 738 (2010).

\bibitem{Glonass}
Glonass Navigation Satellite System, interface controcl document, Edition 5.1, Moscow (2008).

\bibitem{Chine:RSI:2016}
Q. Hao, W. Li, S. He, J. Lv, P. Wang and G. Mei, Rev. Sci. Instr. \bf{87}, \rm 123111 (2016).

\bibitem{Camparo:JAP:1986}
J. C. Camparo, Phys. Today \bf{60}, \rm 33 (2007).

\bibitem{Mileti:IEEE:1998}
G. Mileti, J. Q. Deng, F. L. Walls, D. A. Jennings and R. E. Drullinger, IEEE J. Quantum Electron. \bf{34}, \rm 233 (1998).

\bibitem{Vanier:APB:2007}
J. Vanier and C. Mandache, Appl. Phys. B \bf{87}, \rm 565 (2007).

\bibitem{Micalizio:Metrologia:2012}
S. Micalizio, C. E. Calosso, A. Godone and F. Levi, Metrologia \bf{49}, \rm 425--436 (2012).

\bibitem{Alzetta:PM:1976}
G. Alzetta, A. Gozzini, L. Moi and G. Orriols, Nuovo Cimento B \bf{36}\rm, 5 (1976).

\bibitem{Vanier:APB:2005}
J. Vanier, Appl. Phys. B Lasers Opt. \bf{81}, \rm DOI: 10.1007/s00340-005-1905-3 (2005).

\bibitem{Lin:OL:2012}
J. Lin, J. Deng, Y. Ma, H. He and Y. Wang, Opt. Lett. \bf{37}, \rm 24, 5036--5038 (2012).

\bibitem{Kang:JAP:2015}
S. Kang, M. Gharavipour, C. Affolderbach, F. Gruet and G. Mileti, Journ. Appl. Phys. \bf{117}, \rm 104510 (2015).

\bibitem{Zanon:PRL:2005}
T. Zanon, S. Gu\'{e}randel, E. de Clercq, D. Holleville, N. Dimarcq and A. Clairon, Phys. Rev. Lett. \bf{94}, \rm 193002 (2005).

\bibitem{Danet:UFFC:2014}
J. M. Danet, M. Lours, S. Gu\'erandel and E. De Clercq, IEEE Trans. Ultrason. Ferroelec. Freq. Contr. \bf{61}, \rm 4, 567 (2014).

\bibitem{Dick:PTTI:1987}
G. J. Dick, J. D. Prestage, C. A. Greenhall, and L. Maleki, Local
oscillator induced degradation of medium-term stability in passive
atomic frequency standards, in Proc. 22nd Precise Time and
Time Interval (PTTI) Applications and Planning Meeting, 1990, pp.
487--508.


\bibitem{Yun:APL:2014}
P. Yun, J. M. Danet, D. Holleville, E. De Clercq and S. Gu\'erandel, Appl. Phys. Lett. \bf{105}, \rm 231106 (2014).

\bibitem{Yun:JAP:2016}
P. Yun, S. Gu\'erandel and E. De Clercq, Journ. Appl. Phys. \bf{119}, \rm 244502 (2016).

\bibitem{Yun:PRAp:2017}
P. Yun, F. Tricot, C. E. Calosso, S. Micalizio, B. Fran\c{c}ois, R. Boudot, S. Gu\'erandel and E. De Clercq, Phys. Rev. Applied \bf{7}, \rm 014018 (2017).

\bibitem{Mclocks}
http://www.inrim.it/Mclocks

\bibitem{Jau:PRL:2004}
Y. Y. Jau, E. Miron, A. B. Post, N. N. Kuzma and W. Happer, Phys. Rev. Lett. \bf{93}, \rm 160802 (2004).

\bibitem{Liu:PRA:2013}
X. Liu, J. M. M\'erolla, S. Gu\'{e}randel, C. Gorecki, E. De Clercq and R. Boudot, Phys. Rev. A \bf{87}, \rm 013416 (2013).

\bibitem{MAH:JAP:2015}
M. Abdel Hafiz and R. Boudot, Journ. Appl. Phys. \bf{118}, \rm 124903 (2015).

\bibitem{Castagna:UFFC:2009}
N. Castagna, R. Boudot, S. Gu\'erandel, E. De Clercq, N. Dimarcq, and A. Clairon, IEEE Trans. Ultrason. Ferroelec. Freq. Contr. \bf{56}, \rm 2, 246 (2009).

\bibitem{Boudot:IM:2009}
R. Boudot, S. Gu\'erandel, E. De Clercq, N. Dimarcq and A. Clairon, IEEE Trans. Instr. Meas. \bf{58}, \rm 4, 1217 (2009).

\bibitem{Francois:RSI:2014} B. Fran\c{c}ois, C. E. Calosso, J. M. Danet and R. Boudot, Rev. Sci. Instr. \bf{85}, \rm 094709 (2014).

\bibitem{MAH:OL:2016}
M. Abdel Hafiz, G. Coget, E. De Clercq and R. Boudot, Opt. Lett. \bf{41}, \rm 13, 2982 (2016).

\bibitem{Pitz:PRA:2009}
G. A. Pitz, D. E. Wertepny, and G. P. Perram, Phys. Rev. A \bf{80}, \rm 062718 (2009).

\bibitem{T4science}
http://www.t4science.com/documents/iMaser-Clock-Spec.pdf (iMaser-ST 3000 specifications).

\bibitem{VA}
J. Vanier and C. Audoin, "The quantum physics of atomic frequency standards", Adam Hilger, Bristol (1989).

\bibitem{Zanon:PRA:2011}
T. Zanon-Willette, E. De Clercq and E. Arimondo, Phys. Rev. A \bf{84}, \rm 062502 (2011).

\bibitem{Vanier:PRA:1998}
J. Vanier, A. Godone and F. Levi, Phys. Rev. A \bf{58}, \rm 3, 2345-2358 (1998).



\bibitem{Guerandel:IM:2007}
S. Gu\'erandel, T. Zanon, N. Castagna, F. Dahes, E. De Clercq, N. Dimarcq and A. Clairon, IEEE Trans. Instr. Meas. \bf{56}, \rm 2, 383 (2007).

\bibitem{Gharavipour:FSM:2016}
M. Gharavipour, C. Affolderbach, S. Kang, T. Bandi, F. Gruet, M. Pellaton and G. Mileti, Journal of Physics: Conference Series \bf{723}, \rm 012006 (2016).

\bibitem{Novikova:JOSAB:2016}
E. Kuchina, E. E. Mikhailov and I. Novikova,  Journ. Soc. Am. B \bf{33}, \rm 4, 610-614 (2016).

\bibitem{Yudin:PRA:2010}
V. I. Yudin, A. V. Taichenachev, C.W. Oates, Z.W. Barber,
N. D. Lemke, A. D. Ludlow, U. Sterr, Ch. Lisdat,
and F. Riehle, Phys. Rev. A \bf{82}, \rm 011804(R) (2010).

\bibitem{Miletic:APB:2012}
D. Miletic, C. Affolderbach, M. Hasegawa, R. Boudot, C. Gorecki, and
G. Mileti, Appl. Phys. B \bf{109}, \rm 89-97 (2012).

\bibitem{Huntemann:PRL:2012}
N. Huntemann, B. Lipphardt, M. Okhapkin, Chr. Tamm, E. Peik, A. V. Taichenachev, and V. I. Yudin, Phys. Rev. Lett. \bf{109}, \rm 213002 (2012).

\bibitem{Kozlova:IM:2014}
O. Kozlova, J. M. Danet, S. Gu\'erandel, and E. de Clercq, IEEE Trans. Instrum. Meas. \bf{63}, \rm 1863-1870 (2014).

\bibitem{Yano:PRA:2014}
Y. Yano, W. J. Gao, S. Goka, and M. Kajita, Phys. Rev. A \bf{90}, \rm 013826 (2014).

\bibitem{Pati:JOSAB:2015}
G. S. Pati, Z. Warren, N. Yu, and M. S. Shahriar, J. Opt. Soc. Am. B \bf{32}, \rm 388-394 (2015).

\bibitem{Blanshan:PRA:2015}
E. Blanshan, S. M. Rochester, E. Donley and J. Kitching, Phys. Rev. A \bf{91}, \rm 041401(R) (2015).

\bibitem{Hobson:PRA:2016}
R. Hobson, W. Bowden, S. A. King, P. E. G. Baird,
I. R. Hill, P. Gill, Phys. Rev. A \bf{93}, \rm 010501(R) (2016).

\bibitem{Zanon:Arxiv:2016}
T. Zanon-Willette, E. de Clercq, and E. Arimondo, Phys. Rev. A \bf{93}, \rm 042506 (2016).

\bibitem{Yudin:Arxiv}
V. I. Yudin, A. V. Taichenachev and M. Yu. Basalaev and T. Zanon-Willette, Phys. Rev. A \bf{94}, \rm 052505 (2016).



\bibitem{Borde}
C. Bord\'e, Density matrix equations and diagrams for high-resolution non linear laser spectroscopy: application to
Ramsey fringes in the optical domain, Advances in Laser Spectroscopy, Edited by F. T. Arecchi, F. Strumia and H.
Walther, Plenum Publishing Corporation (1983).










\end{thebibliography}
\end{document}